\documentclass[%
reprint,
superscriptaddress,
 amsmath,amssymb,
 aps,
prb,
floatfix,
]{revtex4-2}

\usepackage[version=3]{mhchem}
\usepackage{graphicx}
\usepackage{dcolumn}
\usepackage{bm}
\usepackage{mathtools}
\usepackage{hyperref}
\usepackage{xcolor}
\usepackage{physics}
\usepackage{array}
\usepackage{multirow}
\usepackage{booktabs}
\usepackage{enumerate}
\usepackage{gensymb}
\usepackage[T1]{fontenc}
\usepackage{amssymb}

\begin{document}


\title{Crystal structure prediction of (quasi-)two-dimensional lead halide perovskites}

\author{Juraj Ov\v{c}ar}
\affiliation{Ru{\dj}er Bo\v{s}kovi\'c Institute, Bijeni\v{c}ka 54, Zagreb, 10000 Croatia}
\author{Luca Grisanti}%
\email{luca.grisanti@irb.hr}
\affiliation{Ru{\dj}er Bo\v{s}kovi\'c Institute, Bijeni\v{c}ka 54, Zagreb, 10000 Croatia}

\author{Bruno Mladineo}%
\affiliation{Ru{\dj}er Bo\v{s}kovi\'c Institute, Bijeni\v{c}ka 54, Zagreb, 10000 Croatia}

\author{Aleksandra B. Djuri\v{s}i\'{c}}%
\affiliation{Department of Physics, The University of Hong Kong, Pokfulam Road, Hong Kong, Hong Kong SAR, China}

\author{Jasminka Popovi\'c}%
\email{jasminka.popovic@irb.hr}
\affiliation{Ru{\dj}er Bo\v{s}kovi\'c Institute, Bijeni\v{c}ka 54, Zagreb, 10000 Croatia}

\author{Ivor Lon\v{c}ari\'c}
\email{ivor.loncaric@gmail.com}
\affiliation{Ru{\dj}er Bo\v{s}kovi\'c Institute, Bijeni\v{c}ka 54, Zagreb, 10000 Croatia}


\begin{abstract}
Two-dimensional lead halide perovskites are promising materials for optoelectronics due to the tunability of their properties with the number of lead halide layers and the choice of an organic spacer. Physical understanding for the rational design of materials primarily requires knowledge of crystal structure. 2D lead halide perovskites are usually prepared in the form of films complicating the experimental determination of structure. To enable theoretical studies of experimentally unresolvable structures as well as high-throughput virtual screening, we present an algorithm for crystal structure prediction of lead halide perovskites. Using automatically prepared classical potential we show that our algorithm enables fast access to a structure that can be used for further first-principles studies.
\end{abstract}

\maketitle


\section{\label{sec:level1}Introduction}
In recent years, metal halide perovskites (MHPs) have proven to be promising candidates for the future materials of choice for manufacturing light-emitting diodes (LEDs) and solar cells.\cite{jena2019halide, lu2019metal} A subclass of MHPs of particular interest for LED applications are the \mbox{(quasi-) 2D} layered perovskites (Q2DPs) due to larger exciton binding energies, improved stability, and wider tunability of properties compared to 3D perovskites.\cite{mao2018two} Particularly, the Ruddlesden-Popper and Dion-Jacobson MHP phases have gained significant attraction.\cite{chen2019merits, ahmad2019dion}
The general chemical formula of Ruddlesden-Popper perovskites (RPPs) is $\text{R}_2\text{A}_{n-1}\text{B}_n\text{X}_{3n+1}$, where $\text{R}^+$ is a large amine spacer cation, $\text{A}^+$ is a smaller organic cation or $\text{Cs}^+$, $\text{B}^{2+}$ is a divalent metal cation, $\text{X}^-$ is a halide anion and $n$ is the number of layers of $\text{BX}_6$ octahedra seperated by a bilayer of $\text{R}^+$ spacer cations (3D perovskite $\text{ABX}_3$ is obtained in the limit $n\rightarrow\infty$). Similarly, the chemical formula of Dion-Jacobson perovskites (DJPs) is $\text{R}\text{A}_{n-1}\text{B}_n\text{X}_{3n+1}$, with the difference compared to RPPs being that $\text{R}^{2+}$ is a \textit{diammonium} cation. This generality of composition offers a great variety of possibilities in choosing $n$ as well as the particular chemicals involved in the synthesis of the layered perovskites, allowing for the characteristic wide tunability of the RPP and DJP physical properties.\cite{mao2018two}
\par
Knowledge of the crystal structure of a material is the starting point for understanding its physical properties. Despite constant advances in methodologies\cite{david2002structure, favre2002fox, david2006dash, altomare2013expo2013}, crystal structure determination from powder diffraction data cannot be yet considered a trivial task because the information from 3D reciprocal space collapses into its 1D projection. Although not as straightforward as the structure solution from single crystals, a huge number of crystalline phases have been successfully solved from the powder diffraction data over the years\cite{cvcerny2017crystal}.  On the other hand, when the sample is prepared in the form of a thin film, as almost always is the case in Q2DP-based LED/solar cell research, the determination of structure becomes close to impossible not only due to the strong influence of crystalline texture on the diffracted intensities but also due to the limited and inadequate methodologies. 
There have been some rare attempts to utilize a grazing incidence X-ray diffraction (GI-XRD) with molecular modeling aiming to determine purely organic structures prepared in the form of thin film\cite{jones2017solution} but on the practical side, the implementation in everyday laboratory work is restricted due to the requirement that data must be collected using synchrotron radiation. Even if the equipment is readily accessible, due to currently underdeveloped methodology that would properly deal with texture-related issues of thin films, such an approach certainly is not efficient enough to elucidate the immense number of novel Q2DP structures that emerge on the daily basis. Considering the inability to solve the structures from the thin film diffraction data, one way to deal with unknown structures in Q2DP films would be to prepare them in the form of a single crystal or powders. 
However, there are experimental difficulties when growing materials in the form of a single crystal\cite{jones2016substrate}, for example, due to limitations in the stability. In this scenario, many structures of large technological potential remain unknown.
\par
Moreover, it would be greatly valuable to know the structure even before the materials are synthesized.
It is, therefore, desirable to use a computational tool to find, understand and predict stable Q2DP crystal structures as well as their physical properties. Different computational approaches have been proposed to explore the possibility of formation of various 3D perovskites\cite{tao2021machine, li2021studies, gomez2021ternary}. Similar computational explorations of Q2DPs have been scarce\cite{jahanbakhshi2021organic, lyu2021predictive} and did not aim to find the global minimum energy structure, most likely due to the prohibitively vast compositional phase space and large system sizes.
\par
Generally, density functional theory (DFT) is utilized abundantly due to its well-known good balance between accuracy and low computational cost. However, in the case of Q2DPs, the approach to crystal structure prediction via DFT suffers from several drawbacks. Optimizing an initial-guess structure using a local structural relaxation algorithm gives no guarantee that the final structure is the global minimum.
This may be resolved by exploring a sufficiently large part of the potential energy surface (PES), but this is computationally unfeasible, as in the case of Q2DPs unit cells may contain several hundreds of atoms, giving rise to complex PES's.\cite{price2020interfacial}
\par
In this work, we introduce a workflow for predicting candidate Q2DP crystal structures using classical model potentials combined with DFT. The potentials are constructed in an automated fashion and then employed to find a global minimum structure via minima hopping algorithm in the vein of Goedecker \cite{goedecker2004minima}, Amsler and Goedecker \cite{amsler2010crystal} and Peterson \cite{peterson2014global}. In this work, we will refer to the developed algorithm as GO-MHALP (Global Optimization via Minima Hopping Algorithm for Layered Perovskites). We aim to develop a method which is generalizable, but specialized to work for Q2DP structures with well defined general structural features, such as the alternating organic/inorganic layered structure. Therefore, we expect to start from structures not too far from a global minimum so that minima-hopping is a well suited technique, as opposed to methods which work well when starting far away from the global minimum, such as particle swarm optimization \cite{wang2012calypso} or genetic algorithms \cite{glass2006uspex}.
Our method works in principle for any candidate Q2DP whose $\text{R}^+/\text{R}^{2+}$ and $\text{A}^+$ compounds are organic cations consisting of $\text{N}$, $\text{C}$ and $\text{H}$ and whose inorganic perovskite octahedra are $\text{Pb}\text{Br}_6$. The complete methodology is described in section \ref{sec:level2} while validations of the model on RPPs containing $\text{BA}$ (butylammonium) and $\text{MA}$ (methylammonium) and a DJP containing $\text{4AMP}$ (4-(aminomethyl)-piperidinium) are presented in section \ref{sec:level3}. Aditionally, we have extended the method to predict an unknown Q2DP structure containing mixed-halide perovskite octahedra $\text{Pb}\text{Br}_3\text{I}_3$ \cite{ovcar2022mixed}.

\section{\label{sec:level2}Methodology}
\begin{figure*}[t]
\includegraphics[width=2\columnwidth]{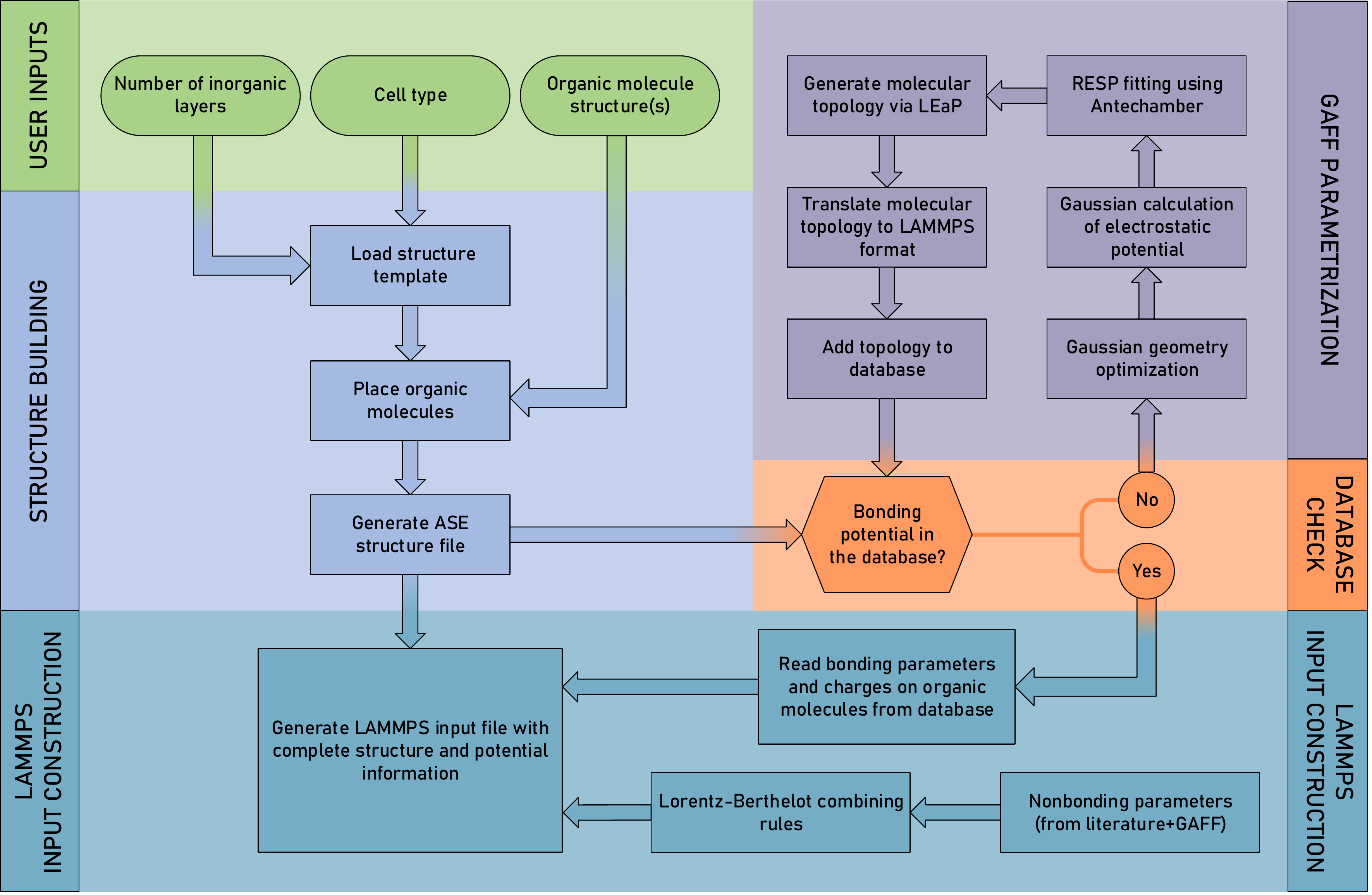}
\caption{Flowchart of the procedure of generating an input LAMMPS file containing information on the structure and interatomic interactions of a Q2DP.}\label{fig:flowchart_potential}
\end{figure*}

\begin{figure*}[t]
\includegraphics[width=2.0\columnwidth]{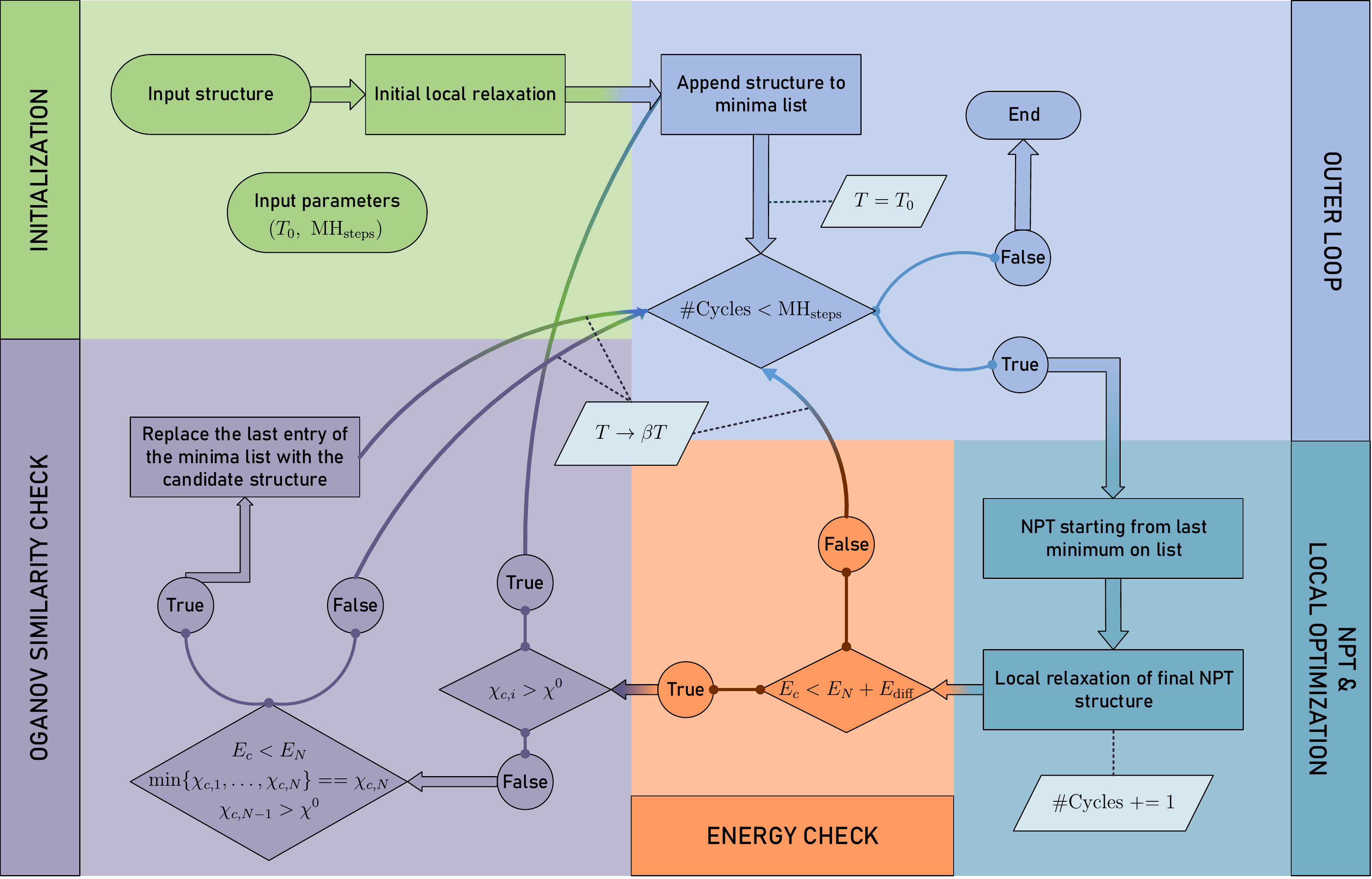}
\caption{Flowchart of the full GO-MHALP algorithm. Dashed lines indicate parameter changes at certain steps.}\label{fig:flowchart_mh}
\end{figure*}

In the following two subsections, we describe the procedure of generating structures and the accompanying classical potentials which are used as inputs for GO-MHALP. A visual aid in the form of a flowchart of the procedure can be viewed in Fig. \ref{fig:flowchart_potential}. In the third subsection, we describe the GO-MHALP algorithm itself with an accompanying flowchart in Fig. \ref{fig:flowchart_mh}.
\subsection{\label{sec:level21}Initial structures generation}
One of the advantages of structure prediction using global optimization algorithms is that the final set of found structures should not depend strongly on the starting structure inputted to the algorithm. However, in the particular case of MH, it is necessary to start from a region in configuration space for which the potential has a physical meaning, i.e. in our case, the starting structure should resemble a Q2DP. To this end, we developed an automatized procedure for the generation of idealized Q2DP structures for a given $\text{R}$ molecule.
\par
\begin{figure}[t]
\includegraphics[width=\columnwidth]{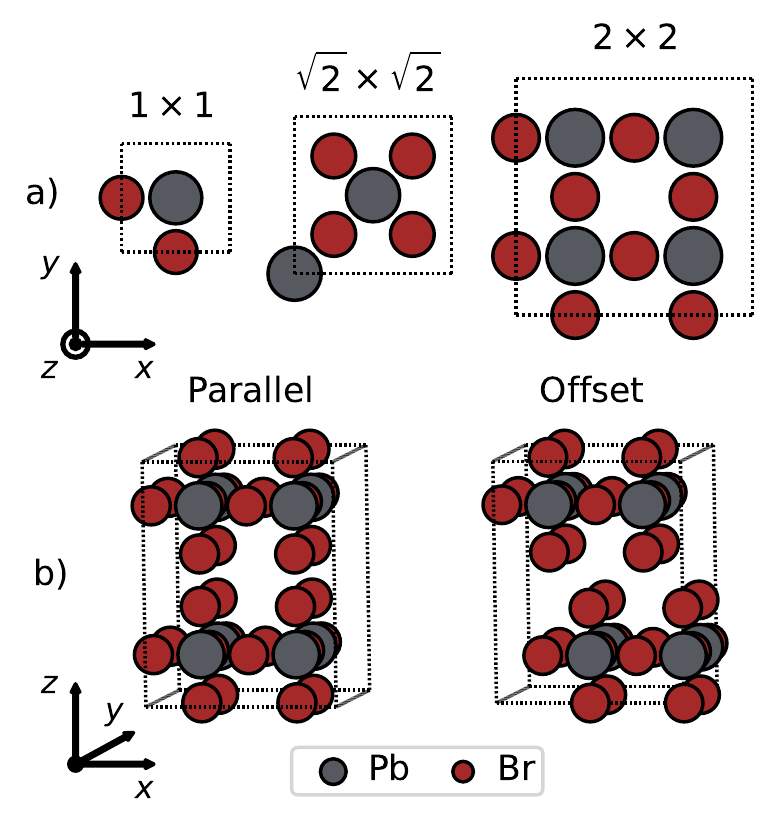}
\caption{Types of initial cell configurations. a) Cross sections perpendicular to the long cell axis with the cell dimensions labelled in units of $\text{Pb}-\text{Pb}$ distance $d_{\text{Pb}-\text{Pb}}=6.080$ \AA. \\b) Interlayer configurations shown for the $2\times 2$ cell type. The long cell axis is shortened and organic molecules are removed for clarity. In the offset configuration the layers are shifted by a quarter of the unit cell length in the $x-$ and $y-$ directions.}\label{fig:structure_types}
\end{figure}

We investigated three types of initially tetragonal cell geometries with parallel or offset interlayer configurations, making a total of six different types of input structures as shown in Fig. \ref{fig:structure_types}. These cell types were chosen because known Q2DP structures often form these geometries\cite{li20212d}. Furthermore, in the various cell types, the supercells can be rearranged to produce equivalent structures, providing an additional check of the independence of the found global minimum on the initial guess structure. The parallel and offset structures are related by a layer shift, while the $1\times 1$ and $\sqrt{2}\times \sqrt{2}$ are subcells of the $2\times 2$ cell type.
\par

We prepared multiple template structures by arranging the inorganic layers in ideal aforementioned configurations and (for $n>1$) placing $\text{MA}$ molecules at the centers of inorganic cages. The $\text{R}^{+}/\text{R}^{2+}$ organic cations are first added to the template structures so that the $\text{NH}_3$ groups of the cations are placed approximately at the centers of the inorganic pockets. For RPPs, the cations are reoriented in such a way so that the vector from the $\text{N}$ atom towards the respective cation's center of mass points in a predefined direction towards the neighboring inorganic sheet. The interlayer spacing between the inorganic sheets can be adjusted as needed to accommodate spacers of various lengths. We advise the reader to see the code in \ref{sec:level5} (Data Availability) as well as Section 3. in Supplemental Material for further details. All generated initial guess structures are given in the Supplemental Material in CIF format as well.
\par
Since we are ultimately interested in global structure optimizations in which the starting structure should not be of decisive importance, this simple way of generating structures works very well for our intentions since it is automatic and fast.
\par
\subsection{\label{sec:level22}Construction of classical potentials}
Following previous work\cite{matsui1987computational, saba2014effect, mattoni2015methylammonium, hata2017development, fridriksson2020tuning}, and in particular 
the idea behind MAPI family of potential developed by Mattoni et al. \cite{mattoni2015methylammonium,hata2017development},
we undertake an approach where the total classical potential is modeled as a sum of i) nonbonding potential that depends only on the interatomic distances with only two-body terms taken into account and ii) a bonding potential including bonds, dihedrals, and angles as described by GAFF\cite{wang2004development}, a generalization of the AMBER\cite{ponder2003force} force field. 

Denoting the list of positions of the nuclei with $\{\boldsymbol{R}\}\coloneqq (\boldsymbol{R}_1,\boldsymbol{R}_2,\ ...\ ,\boldsymbol{R}_N)$, the general form of the total potential energy can be written as
\begin{align}\label{eqn:general_potential}
    U(\{\boldsymbol{R}\}) = \frac{1}{2}\sum_{i,j}^N U_{ij}(R_{ij}) + U^{\text{bonding}},
\end{align}
where $R_{ij}\coloneqq\abs{\boldsymbol{R}_i-\boldsymbol{R}_j}$. 
As in all effective-potential approaches, the electronic degrees of freedom do not enter the calculation explicitly but are rather absorbed into the effective interactions of the nuclei.\cite{marx2009ab}\par
$U_{ij}$ is separated into three parts, dealing with the nonbonding inorganic-inorganic, inorganic-organic and organic-organic interatomic interactions respectively:
\begin{align}\label{eqn:U_seperation}
    U_{ij} = U_{ij}^{\text{II}} + U_{ij}^{\text{IO}} + U_{ij}^{\text{OO}}.
\end{align}
The explicit form of the inorganic-inorganic interaction is
\begin{align}\label{eqn:U_II}
    U_{ij}^{\text{II}}(R_{ij}) = A_{ij}\exp (-R_{ij}/\rho_{ij}) - \frac{c_{ij}}{R_{ij}^6} + \frac{ q_iq_j}{4\pi\epsilon_0R_{ij}},
\end{align}
where the first two terms comprise the Buckingham potential\cite{buckingham1938classical} with the first term describing the Pauli repulsion at small nuclei distance and the second term describing the van der Waals interaction with $A_{ij}, \rho_{ij}$ and $c_{ij}$ as model parameters. The final term is the Coulomb interaction between two (possibly partial) ionic charges. The inorganic-organic interaction is modelled as follows:
\begin{equation}\label{eqn:U_IO}
\begin{split}
    U_{ij}^{\text{IO}}(R_{ij})&=A_{ij}\exp (-R_{ij}/\rho_{ij}) - \frac{c_{ij}}{R_{ij}^6} + \frac{ q_iq_j}{4\pi\epsilon_0R_{ij}}\\
    &+4\epsilon_{ij}\Big[-\Big(\frac{\sigma_{ij}} {R_{ij}}\Big)^6 + \Big(\frac{\sigma_{ij}} {R_{ij}}\Big)^{12}\Big],
\end{split}
\end{equation}
i.e., besides the Buckingham and Coulomb terms a Lennard-Jones term is added with additional parameters $\epsilon_{ij}$ and $\sigma_{ij}$. For $(\text{Pb, Br})\text{ - }(\text{C, N})$ interactions only the Buckingham and Coulomb terms are used, while only Coulomb and Lennard-Jones terms are used for $(\text{Pb, Br})\text{ - }\text{H}$ interactions. Similarly, nonbonding organic-organic interactions are described only by Lennard-Jones and Coulomb terms:
\begin{align}\label{eqn:U_OO}
    U_{ij}^{\text{OO}}(R_{ij}) = 4\epsilon_{ij}\Big[-\Big(\frac{\sigma_{ij}} {R_{ij}}\Big)^6 + \Big(\frac{\sigma_{ij}} {R_{ij}}\Big)^{12}\Big] + \frac{ q_iq_j}{4\pi\epsilon_0R_{ij}}.
\end{align}
For $U_{ij}^{II}$, $U_{ij}^{IO}$ and $U_{ij}^{OO}$ a cutoff parameter $r_c$ is used so that for $R_{ij} > r_c$ only the long range Coulomb interaction is calculated using the $\text{P}^3\text{M}$ algorithm.\cite{hockney1988computer_kspacepppm} The bonding potential which, of course, concerns only interactions within organic molecules, has the following form:
\begin{equation}\label{eqn:Ubonding}
    \begin{split}
        U^{\text{bonding}} &= \sum_{ij}^{N} K^\text{b}_{ij}(R_{ij} - R^0_{ij})^2 \\
        &+ \sum_{ijk}^{N} K^\text{a}_{ijk}(\theta_{ijk}-\theta^0_{ijk})^2 \\
        &+ \sum_{ijkl}^{N} K^\text{d}_{ijkl}\big(1+\cos(n_{ijkl}\phi_{ijkl}-\phi^0_{ijkl})\big)
    \end{split}
\end{equation}

expressing bonds, angles and dihedrals respectively. $R_{ij}$ are two-body distances, $\theta_{ijk}$ are three-body angles and $\phi_{ijkl}$ are four-body-dihedrals while the other factors are GAFF parameters. The main advantage of the GAFF force field is its capability to describe a very large number of organic molecules with an acceptable level of accuracy. Besides, its standard working frame allows in principle an automatic atom type assignment for any given organic molecules in its (reliably) relaxed geometry.
\par
The idea of GO-MHALP is to have a general and flexible tool that can be applied to any organic molecule $\text{R}$ in a candidate Q2DP structure. 
If the particular molecule is not found in the local database of organic cations the potential can be generated in the very first step provided a starting geometry along with the GAFF philosophy.
Shortly, to obtain GAFF parameters the geometry of the $\text{R}^+$ molecule is optimized using the Gaussian\cite{g09} program with the B3LYP\cite{becke1988density, lee1988development, vosko1980accurate, stephens1994ab} hybrid DFT functional and 6-311G* basis set \cite{mclean1980contracted, clark1983a, krishnan1980a}. 
Consistently with previous work \cite{mattoni2015methylammonium}, 
the electrostatic potential (ESP) of the optimized isolated cation with $+1$ total charge is obtained via the BP86\cite{becke1988density, becke1993density, perdew1986accurate, perdew1992accurate} GGA functional and the Def2TZVP\cite{weigend2005balanced, weigend2006accurate} basis set.
Partial atomic charges are then obtained by directly fitting this \textit{ab initio} electrostatic potential (ESP) using the restrained electrostatic potential (RESP)\cite{bayly1993well} method as implemented in the Antechamber\cite{wang2006automatic} program from the Amber16 suite\cite{duke2016amber, salomon2013overview}. Complete molecular topology and GAFF parameters are generated and translated to LAMMPS\cite{plimpton1995fast} format and added to the local database of organic cations.
\par
The parameters for the $\text{C}-\text{C}$, $\text{N}-\text{N}$ and $\text{H}-\text{H}$ nonbonding Lennard-Jones interactions are obtained from GAFF parametrization as described above. As for the nonbonding parameters concerning $\text{Pb}-\text{Pb}$ and $\text{Br}-\text{Br}$ interactions, the values of these are taken to be the same as the ones obtained in Hata \textit{et al.}\cite{hata2017development} for $\text{MAPbBr}_3$, barring the charges $q_i$, which for organic molecules we set to the partial charges obtained by RESP fitting and for the atoms comprising the inorganic perovskite structure we set $q_{\text{Pb}}=+2$ and $q_{\text{Br}}=-1$. The reason for this is that the charges in Hata \textit{et al.} were set by rescaling of charges obtained by Mattoni \textit{et al.}\cite{mattoni2015methylammonium} for $\text{MAPbI}_3$ and in both of these works the obtained charges of the inorganic lattice $\text{PbBr}_3$ do not sum up to 1. In Mattoni et al. system neutrality was then ensured by refitting the charge parameters of the whole model (including $\text{MA}$) to data obtained via DFT. For this work, by setting integer charges we avoid the need for expensive DFT calculations and refitting procedures, thus greatly increasing the transferability of this method for construction of classical potentials.
\par A benchmark of the accuracy of the potentials may be found in the Supplemental Material.
\subsection{\label{sec:level23}Minima hopping}
Minima hopping (MH) is an efficient and simple global optimization method first developed by Goedecker\cite{goedecker2004minima}. The general idea is to alternate between molecular dynamics (MD) simulations and local structure optimizations (relaxations) after which some criteria are used to determine whether the optimized structure will be accepted as a newfound local minimum. A system may gain enough kinetic energy during MD to overcome a potential barrier and in this way, a complex PES may be traversed to arrive at different potential energy basins.\par
The original MH method conceived by Goedecker concerned only nonperiodic systems and therefore employed only local optimizations of atomic positions. However, the ground state of a crystalline system is fully determined not only by atomic positions but also by the unit cell parameters. Therefore, by using variable cell shape MD, Amsler and Goedecker generalized the MH method to be functional for periodic systems as well.\cite{amsler2010crystal} Another variant of MH, dubbed constrained minima hopping, was developed by Peterson\cite{peterson2014global}. Peterson introduced a simple constraint, based on Hooke's law, in order to prevent dissociation of molecules adsorbed on a surface during the MD portion of MH, thereby effectively reducing the configurational space to be explored by MH only to configurations of interest (namely, those where the adsorbate identity is preserved).\par
In this work, we combine and modify the approaches introduced above to develop an MH algorithm suitable for Q2DP global structure optimization: GO-MHALP. We simulate MD in the isothermal-isobaric (NPT) ensemble and optimize the unit cell as well as atomic positions during local optimizations. Additionally, we developed a new scheme for detailed exploration of local energy basins and on-the-fly selection of the lowest energy structures found in them.
\par
We implemented our algorithm based on an MH algorithm already existing within the ASE\cite{larsen2017atomic} package using the LAMMPSlib interface to read in the classical potential described in section \ref{sec:level22}. Now we present an outline of the algorithm. A list of initial parameters, their descriptions and values can found in Table \ref{table:MH_pars}.
\par
The input structure is fed into the algorithm and its cell and atomic positions are optimized. This optimized structure is the first entry in the list of found local minima. The initial optimization is performed with a looser force convergence threshold than following optimizations to avoid a long optimization step since the initial structure may be far away from a local minimum. For ease of writing, from now on we will label the physical properties of the entries in the list of found local minima with the subscript $i$, where $i=1,\dotsc ,N$ so that $N$ labels the last found minimum. At this point, there is only one structure in the list of found minima, i.e. $N=1$.
\par
An NPT\cite{melchionna1993hoover, melchionna2000constrained, holian1990time} molecular dynamics simulation is performed starting from the locally optimized structure at temperature $T=T_0$ and with other parameter values being as listed in Table \ref{table:MH_pars}. The values of the NPT related parameters were chosen to ensure that the molecular dynamics is long enough for the system to completely thermalize. Generally, the starting NPT configuration is the last
($N$-th) minimum in the list of found minima and $T$ varies during GO-MHALP as described below. Ions are given random initial velocities corresponding to a Maxwell-Boltzmann distribution of temperature $T$. NPT is stopped after $\text{md}_\text{min}$ local minima have been passed over with one pass counted if a sequence of potential energies calculated at each MD step ends with two downward points followed by two upward points.
\par
The atomic positions and cell parameters of the last configuration obtained by NPT dynamics are optimized. We label the physical properties of this candidate structure with the subscript $c$. After optimization, energy and structure similarity checks are performed to determine whether the candidate structure will be added to the list of local minima. Firstly, $E_c$ is compared to $E_N$; if $E_c>E_N+E_{\text{diff}}$, the structure is discarded as being too high in energy, NPT temperature is increased, i.e. $T\rightarrow\beta T$ and a new NPT/optimization cycle is started from the $N$-th minimum in the list of found minima. Otherwise, a structure similarity check is performed. For structure comparison we use Oganov fingerprints\cite{oganov2009quantify, lyakhov2010predict} which were first used in the context of MH by Amsler and Goedecker\cite{amsler2010crystal}. With this method, for every structure a unique "fingerprint" may be calculated and represented as a matrix in an abstract vector space. The components of the matrix are sums of Gaussian-smeared delta functions:
\begin{align}
    F_{AB}(R)=\sum_{A_k,B_l}\frac{\delta(R-R_{kl})}{4\pi R_{kl}^2\frac{N_A N_B}{V}\Delta}-1,
\end{align}
where the sum runs over all pairs of atoms of types $A$ and $B$ found within the cutoff distance $R_{kl}<R_c$, with $N_A$ and $N_B$ being the number of atoms of the respective type $A$ and $B$ in the unit cell and $V$ being the unit cell volume. Each fingerprint component $F_{AB}(R)$ is discretized into bins of width $\Delta$ so it can be represented as a vector with the $m$-th vector coordinate being the value of the fingerprint component $F_{AB}(m)$. A \textit{cosine distance} $\chi_{i,j}$ may then be defined as a measure of dissimilarity of structures $i$ and $j$:
\begin{align}
 \chi_{i,j}=\frac{1}{2}\Bigg(1-\frac{\sum_{AB}\sum_{m}F^i_{AB}(m)F^j_{AB}(m)w_{AB}}{\sqrt{W^i W^j}} \Bigg)   
\end{align}
where the importance weight $w_{AB}$ is defined as
\begin{align}
    w_{AB}=\frac{N_A N_B}{\sum_{\text{cell}}N_A N_B}
\end{align}
and $W^i$ is the norm
\begin{align}
    W^i=\sum_{AB}\sum_{m}\Big[F^i_{AB}(m)\Big]^2w_{AB}.
\end{align}
To reduce noise, we excluded hydrogen atoms from the calculation of cosine distances. The cosine distance $\chi_{c,i}$ is calculated as a measure of dissimilarity between the candidate structure and a structure $i$ from the list of found local minima. If $\chi_{c,i}>\chi^0$ for every $i$ and the threshold parameter $\chi^0$, the candidate structure is added to the local minima list and the NPT temperature is reset to the initial temperature $T_0$. Otherwise, the candidate structure is considered not to be a unique minimum and the NPT temperature is increased, i.e. $T\rightarrow\beta T$. The candidate structure replaces the $N$-th minimum if the following three conditions are met: $E_c<E_N$, $\min\{\chi_{c,1},\dotsc,\chi_{c,N}\}=\chi_{c,N}$ and $\chi_{c,N-1}>\chi^0$. With this replacement scheme and the choice of parameters as listed in Table \ref{table:MH_pars}, we found that the algorithm correctly explores local potential energy basins while preserving structural dissimilarity of the found minima.
\par
This concludes a complete MH cycle in our GO-MHALP frame. If the number of cycles is less than the given $\text{M}\text{H}_\text{steps}$ parameter the algorithm will start a new NPT simulation from the last found minimum and otherwise the algorithm stops.
\subsection{\label{sec:level24}DFT calculations}
On specific structures (see below), DFT relaxations were additionally performed in order to validate and refine the results. All DFT relaxations were performed using the plane-wave basis set code Quantum Espresso \cite{giannozzi2009quantum, giannozzi2017advanced} with the plane-wave basis set cutoff being 816 eV. GBRV pseudopotentials \cite{garrity2014pseudopotentials} were employed together with the  vdW-DF-cx exchange-correlation functional \cite{hyldgaard2015van}. A Monkhorst-Pack k-point mesh \cite{monkhorst1976special} with a density of 5 $\text{\AA}$ was used for Brillouin zone integration. The atomic positions and the unit cell parameters were relaxed until the pressure, the forces on each atom and the total energy change were smaller than 0.5 kbar, 0.02 eV $\text{\AA}^{-1}$  and 1 meV, respectively. 
\subsection{\label{subsec:level24}Similarity measures of simulated powder XRD patterns}
\par
\begin{figure}[t]
\includegraphics[width=1\columnwidth]{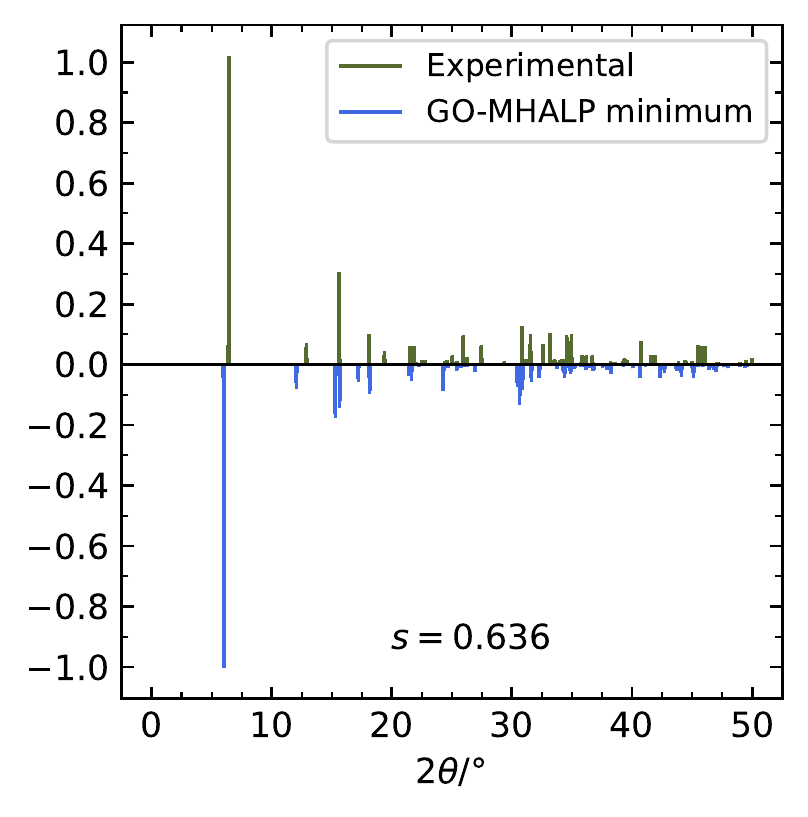}
\caption{Comparison of simulated PXRDs of experimentally solved and one of the structures predicted with GO-MHALP of $\text{BA}_2\text{Pb}\text{Br}_{4}$. PXRD intensity is scaled so that the highest peak has the value $1.0$. In this example the similarity measure amounts to $s=$0.636.}\label{fig:pxrd_example}
\end{figure}

In order to assess the validity of the predicted structures with our protocol, we quantified the similarity of structures obtained with GO-MHALP to structures solved from single-crystal XRD data by simulating their powder XRD (PXRD) patterns and calculating a similarity measure based on cross-correlation functions\cite{de2001generalized, habermehl2014structure} as implemented in the PyXtal\cite{fredericks2021pyxtal} Python library. Explicitly, the similarity measure $s_{12}$ of two powder diagrams $y_1(\theta)$ and $y_2(\theta)$, invariant against scaling of the PXRD intensities, is calculated as
\begin{align}
    s_{12}=\frac{\int w(r)c_{12}(r)dr}{\Big[\int w(r)c_{11}dr\int w(r)c_{22}(r)dr\Big]^{1/2}}
\end{align}
where $c_{12}$ is the cross-correlation function:
\begin{align}
    c_{12}(r)=\int y_1(\theta)y_2(\theta+r)d\theta
\end{align}
with the auto-correlation functions $c_{11}$ and $c_{22}$ defined analogously. We used the cosine weighting function:
\begin{align}
    w(r)=
    \begin{cases}
    0.5\big(\cos(\pi\frac{r}{l})+1\big),&\quad\abs{r}<l \\ 
    0,&\quad\abs{r}>l
    \end{cases}
\end{align}
with the cutoff $l=1.0\degree$. The similarity measure adopts values between 0 and 1, where $s_{12}=1$ corresponds to identical PXRDs. An example of a comparison of PXRDs simulated from an experimentally solved structure and a minimum obtained with GO-MHALP is shown in Figure \ref{fig:pxrd_example}.
\subsection{\label{subsec:level25}Potentials for iodine and mixed halide systems}
GO-MHALP is a general procedure which can in principle, given a suitable classical potential, be used for any systems. E.g., classical potentials for Q2DPs containing iodine instead of bromide can be constructed in a completely analogous manner as described in Section \ref{sec:level22}, but using the parameters for $\text{MAPbI}_3$.\cite{mattoni2015methylammonium, fridriksson2020tuning}. As detailed in an another work\cite{ovcar2022mixed}, we employed GO-MHALP to predict a mixed halide structure. We use the Berthelot rule to calculate the Buckingham parameters for $\text{Pb}-\text{Pb}$ and $\text{Br}-\text{I}$ interactions:
\begin{align}
    A^{\text{mixed}}=\sqrt{A^{\text{Br}}A^{\text{I}}},
\end{align}
where $A^{\text{Br}}$ and $A^{\text{I}}$ denote Buckingham parameters used for pure halide structures and $A^{\text{mixed}}$ are Buckingham parameters used for the mixed halide structures.

\section{\label{sec:level3}Results and discussion}

We first validate GO-MHALP on the well known case of $\text{R}^+=\text{BA}^+$ (butylammonium) cation as spacer. Both $\text{BA}_2\text{Pb}\text{Br}_4 $ and $\text{BA}_2\text{MA}\text{Pb}_2\text{Br}_{7}$ have been successfully prepared and their crystal structures were solved.\cite{gong2018electron,li2019two} Following the tests on RPPs with $\text{BA}$ we continue the validation of GO-MHALP on a DJP structure containing $\text{4AMP}$ (4-(aminomethyl)-piperidinium).\cite{mao2018two} We use the experimentally obtained structures as reference points for validation of GO-MHALP predictions. Radial distribution functions and simulated PXRD patterns for relevant (predicted and experimental) structures may be found in the Supplementary Information. CIF files for these structures are given the Supplementary Data.
\par
Finally, we show performance of GO-MHALP to predict a structure of a mixed-halide Q2DP $\text{\textit{t}-BA}_2\text{Pb}\text{Br}_{2}\text{I}_{2}$. This structure was experimentally solved after the prediction with GO-MHALP.

\begin{figure}[!ht]
\fbox{\includegraphics[width=0.8\columnwidth]{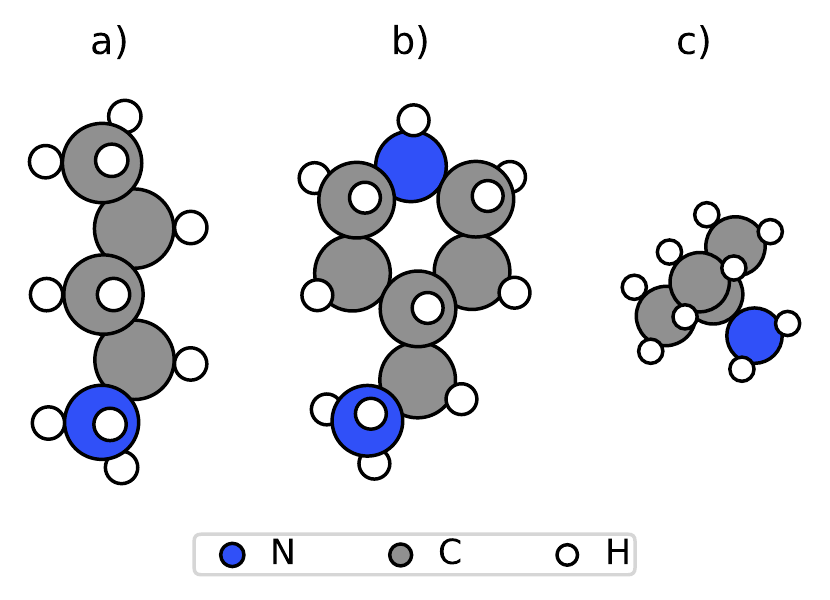}}
\caption{Organic spacers for which the Q2DP structure was predicted with GO-MHALP: a) butylammonium ($\text{BA}$), b) 4-(aminomethyl)-piperidinium ($\text{4AMP}$), c) \textit{tert}-butyl ammonium ($\text{\textit{t}-BA})$.} \label{fig:molecules}
\end{figure}
\begin{figure}[!ht]
\includegraphics[width=0.8\columnwidth]{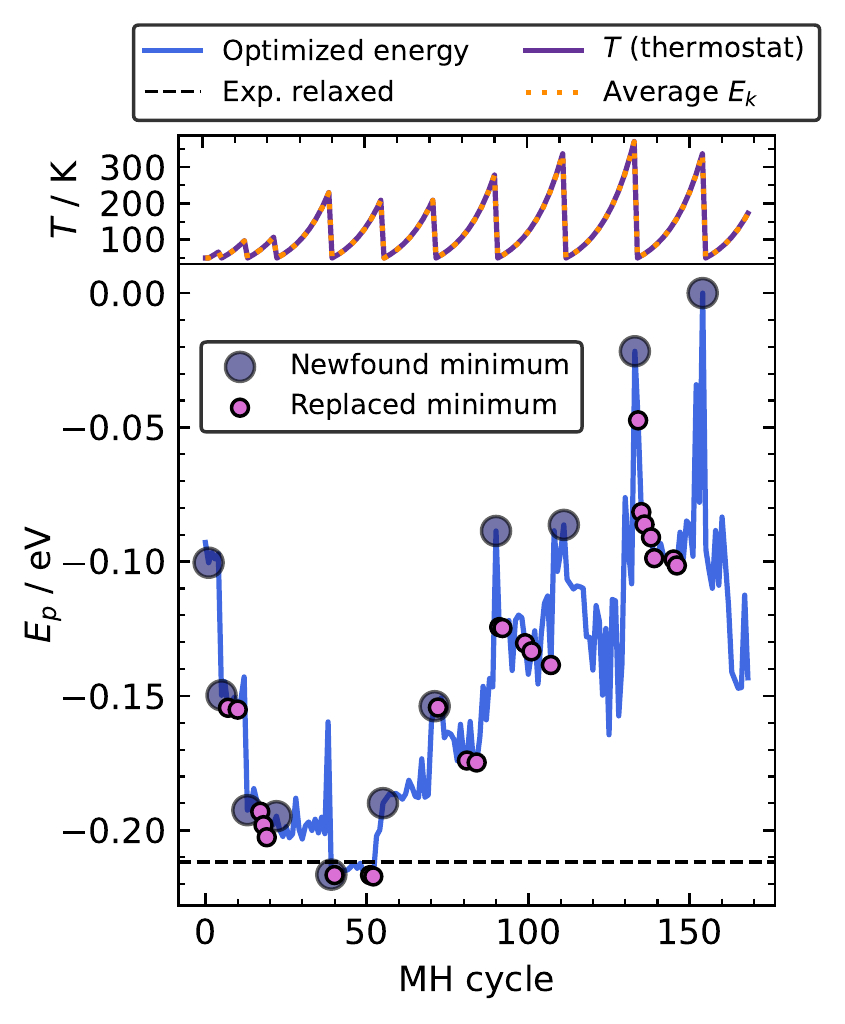}
\caption{Summary of a GO-MHALP run for $\text{BA}_2\text{Pb}\text{Br}_{4}$ with the $2\times 2$ cell type and initially offset layers. Top panel: NPT thermostat temperature and the average kinetic energy of the last $20$ ps of MD across MH cycles. Bottom panel: 
classical potential energies of the candidate structures across the run (blue) and the experimentally solved structure relaxed with the classical potential (dashed black). Identified local minima are marked with circles. }\label{fig:ban1_mh}
\end{figure}
\begin{figure*}[!ht]
\includegraphics[width=1.6\columnwidth]{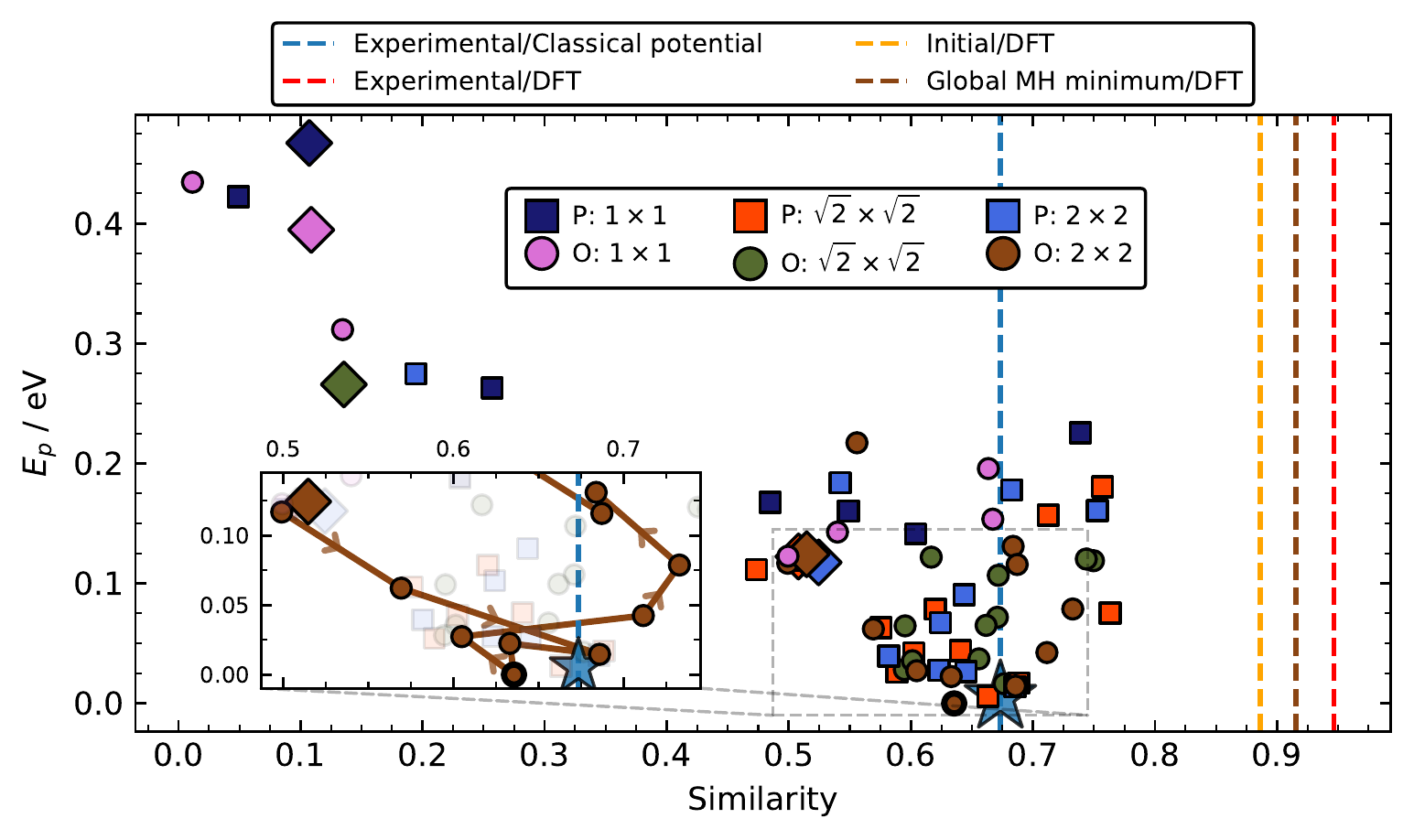}
\caption{Similarity measures and potential energies of minima obtained with GO-MHALP for $\text{BA}_2\text{Pb}\text{Br}_{4}$.
The vertical dashed lines labelled in legend as X/Y show similarity measures of structures X relaxed with Y level of theory (see text). Different markers indicate different cell types, where P(O) indicates an initially parallel (offset) structure (see section \ref{sec:level21}). Diamond $(\blacklozenge)$ markers indicate the starting GO-MHALP points after initial relaxation. The star$(\bigstar)$ marker indicates the similarity measure and energy of the experimental structure relaxed with the classical potential. The order of the accepted minima for the optimal cell type is shown in the inset with the global GO-MHALP minimum emphasized with a bolded edge. Energies are shown per formula unit with the zero of the potential energy chosen as the energy of the global GO-MHALP minimum.}\label{fig:similarity_vs_en_ban1}
\end{figure*}
\begin{figure}[!ht]
\includegraphics[width=0.8\columnwidth]{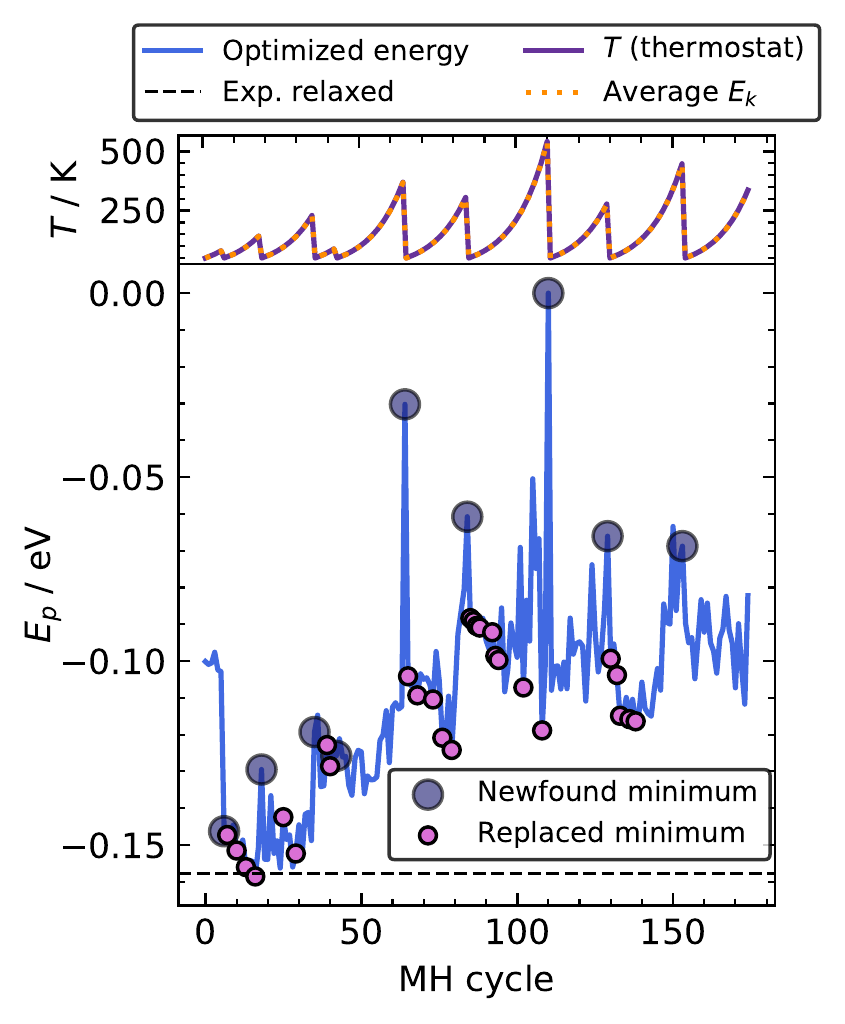}
\caption{Summary of a GO-MHALP run for $\text{BA}_2\text{MA}\text{Pb}_2\text{Br}_{7}$ with the $\sqrt{2}\times \sqrt{2}$ cell type and initially parallel interlayer configuration. 
Top panel: NPT thermostat temperature and the average kinetic energy of the last $20$ ps of MD across MH cycles. Bottom panel: 
classical potential energies of the candidate structures across the run (blue) and the experimentally solved structure relaxed with the classical potential (dashed black). Identified local minima are marked with circles.}\label{fig:ban2_mh}
\end{figure}
\begin{figure}[!ht]
\includegraphics[width=0.8\columnwidth]{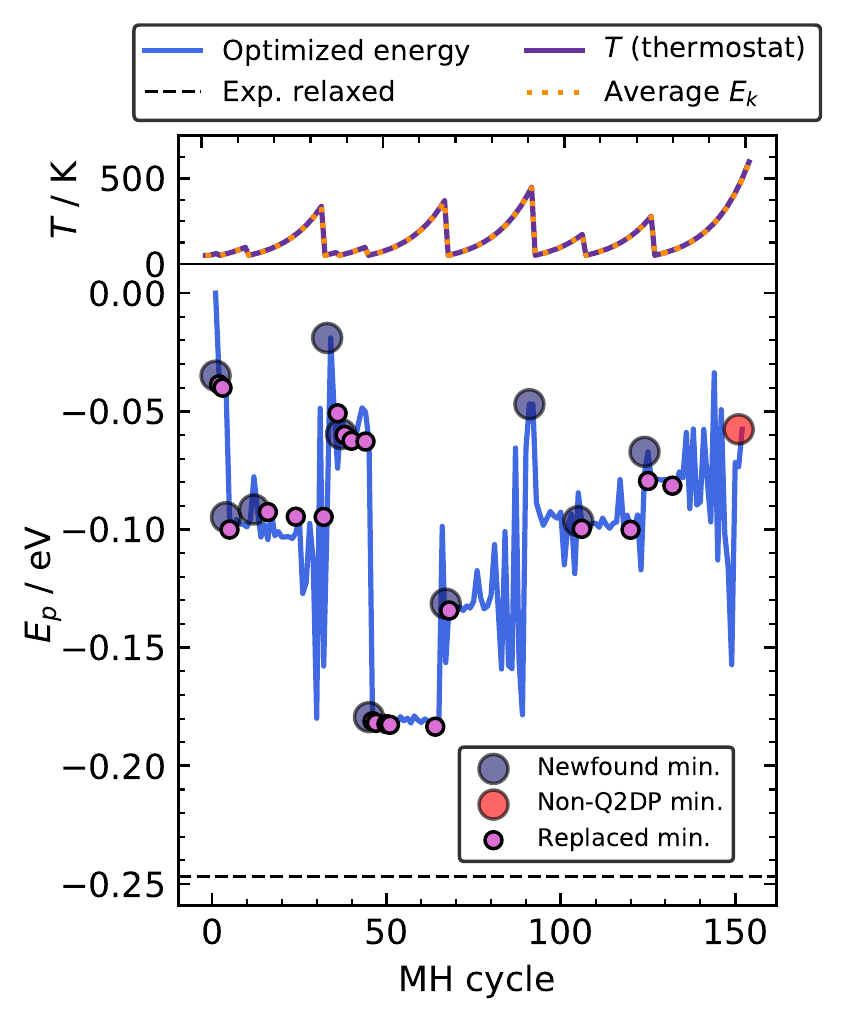}
\caption{Summary of a GO-MHALP run for $\text{(4AMP)}\text{Pb}\text{Br}_{4}$ with the $\sqrt{2}\times \sqrt{2}$ cell type, initially parallel interlayer configuration and alternating $\text{4AMP}$ molecules configuration. 
Top panel: NPT thermostat temperature and the average kinetic energy of the last $20$ ps of MD across MH cycles. Bottom panel: 
classical potential energies of the candidate structures across the run (blue) and the experimentally solved structure relaxed with the classical potential (dashed black). Identified local minima are marked with circles.}\label{fig:4amp_mh}
\end{figure}
\begin{figure*}[!ht]
\includegraphics[width=1.6\columnwidth]{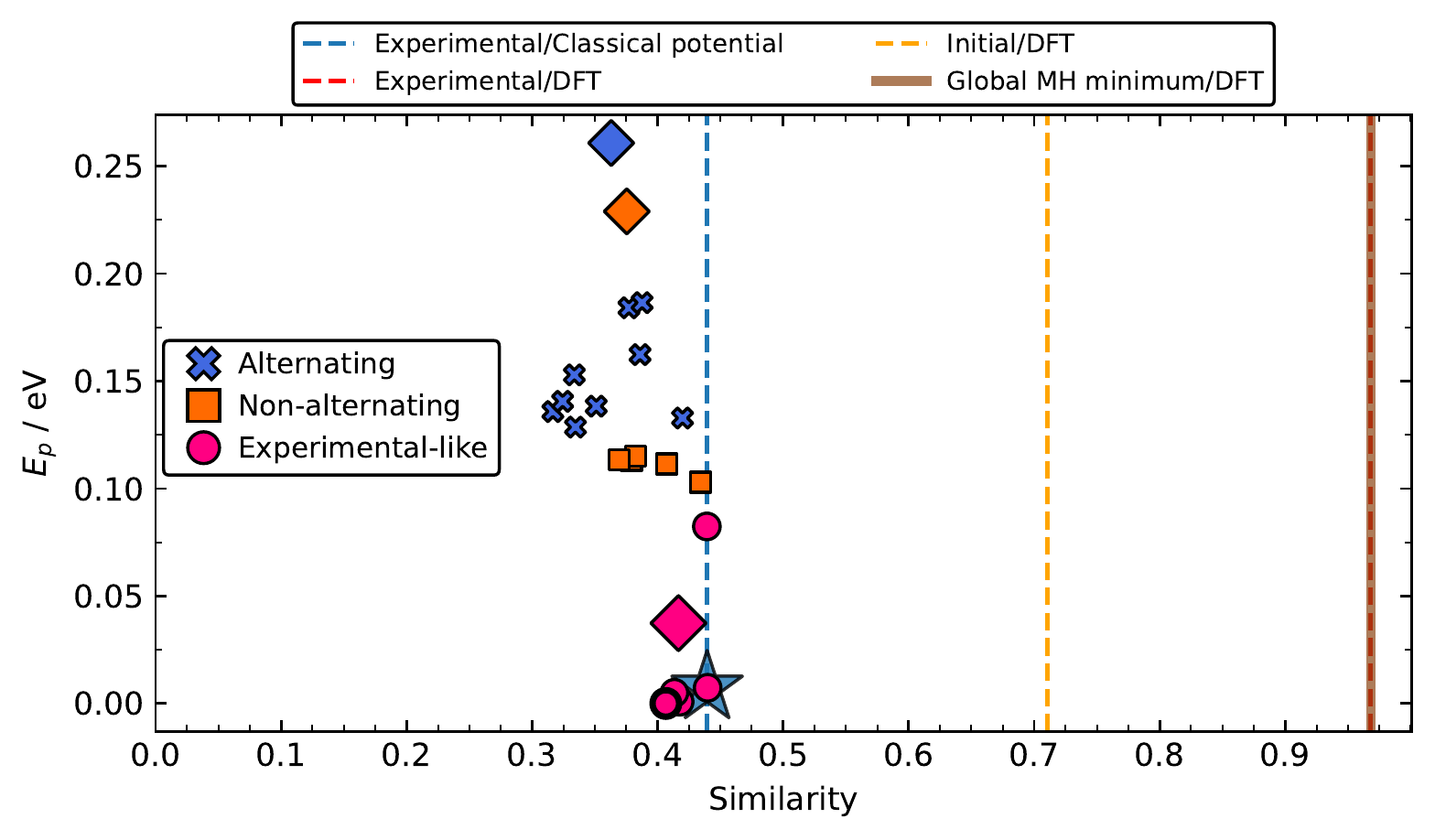}
\caption{Similarity measures and (classical) potential energies of $\text{(4AMP)}\text{Pb}\text{Br}_{4}$ starting from the experimental-like $2\sqrt{2}\times\sqrt{2}$ cell type. The vertical dashed lines indicated as X/Y show similarity measures of structures X relaxed with Y level of theory (see text). Different markers indicate different patterns of connection of $\text{4AMP}$ molecules to the inorganic perovskite layer (see text). Diamond $(\blacklozenge)$ markers indicate the starting GO-MHALP points after initial relaxation. The star $(\bigstar)$ marker indicates the similarity measure and energy of the experimental structure relaxed with the classical potential. Energies are shown per formula unit with the zero of the potential energy chosen as the energy of the global GO-MHALP minimum.}\label{fig:similarity_vs_en_4amp_explike}
\end{figure*}
\begin{figure}[!ht]
\includegraphics[width=1\columnwidth]{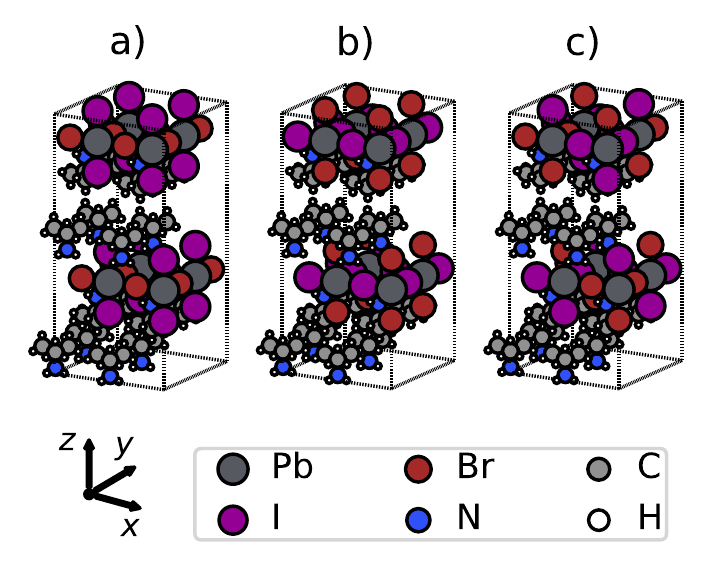}
\caption{Starting GO-MHALP structures for $\text{\textit{t}-BA}_2\text{Pb}\text{Br}_{2}\text{I}_{2}$. The cell type is $2\times 2$. The vectors from the nitrogen atoms of the $\text{\textit{t}-BA}$ molecules towards the respective centers of mass are aligned with the $z$-axis. The structures differ in the halide distrubution, with the bromides (iodides) occupying a) equatorial (axial), b) axial (equatorial) and c) alternating positions.}\label{fig:tba_starting_structures}
\end{figure}
\subsection{\label{subsec:level31}$\text{BA}_2\text{Pb}\text{Br}_{4}$}
We first applied GO-MHALP to $\text{BA}_2\text{Pb}\text{Br}_{4}$ RPP. The six types of unit cells described in \ref{sec:level21} were used as inputs. A summary of a GO-MHALP run is shown in Fig. \ref{fig:ban1_mh}. First, the top panel shows that during MD the system is well thermalized to the set thermostat temperature. Potential energy barriers are overcame by gradually increasing the temperature and newfound minima are accepted upon entrance into a local energy basin. After each restart of the NPT temperature to $T_0$, GO-MHALP explores the surrounding configuration space in detail and the replacement scheme described in section \ref{sec:level23} selects the lowest energy structure found in a basin. 
The global minimum for this run, i.e. the lowest energy local minimum, is found in a distinct basin that lies very close in energy to the experimentally solved structure whose atomic positions and cell parameters were optimized with the classical potential (dashed line). This is an indication that the global minimum of the model potential is connected to the true (experimental) global minimum by a local optimization, a point to which we will return below.
\par
For a complete test and validation of GO-MHALP, we calculated similarity measures of PXRDs between the experimental structure and
\begin{enumerate}[a]
    \item[a)] each of the final minima predicted with six GO-MHALP runs for the six different cell types;
    \item[b)] experimental structure, relaxed with the classical potential;
    \item[c)] experimental structure, relaxed with DFT;
    \item[d)] lowest energy (global) minimum found with GO-MHALP, relaxed with DFT;
    \item[e)] the GO-MHALP initial
    structure from which the global minimum was found, relaxed with DFT.
\end{enumerate}
Similarity measures a) and b) are plotted against classical potential energies in Fig. \ref{fig:similarity_vs_en_ban1} as scatter points and similarity measures b)-e) are plotted as vertical dashed lines. In particular, the line corresponding to case (b) sets the practical limit of the similarity measure that can be reached by this version of GO-MHALP.
\par
First of all, we can notice that the energies of the best (lowest energy) minima of $1\times 1$ cell types are noticeably higher than the larger cell types, meaning that GO-MHALP predicts that $1\times 1$ unit cells are too small to capture all experimentally realized degrees of freedom, which is indeed correct as the experimental structure is of the $\sqrt{2}\times\sqrt{2}$ type. The best minima of $\sqrt{2}\times\sqrt{2}$ and $2\times 2$ cell types cluster nearby the experimental structure relaxed with the classical potential (marked with a star) regardless of the initially parallel or offset interlayers, showing that in these cases GO-MHALP reliably finds the global minimum of the potential regardless of the details of the input structures.
While GO-MHALP finds the experimental structure relaxed with the classical potential, agreement of these structures with the true experimental structure is not completely satisfactory (similarity of around 0.7). On the other hand, relaxing the experimental structure with DFT achieves a similarity of 0.95.
Clearly, GO-MHALP has the ability to find the global minimum, but the model potential should be improved.
\par
The global minimum itself is found for the initially O: $2\times 2$ cell type, with the best P: $2\times 2$ and the $\sqrt{2}\times \sqrt{2}$ minima being slightly higher in energy. O: $2\times 2$ input structure after initial relaxation shows already a good similarity of $0.515$ which is further improved by GO-MHALP to score $0.635$ at the global minimum. 
Relaxing the global minimum with DFT significantly improves this value to a similarity measure of $0.915$. This final step suggests that an extra DFT relaxation of the global minimum found by GO-MHALP renders final structures that can be highly accurate. 
The small inset in Figure \ref{fig:similarity_vs_en_ban1} depicts how the PES exploration works in GO-MHALP in the case of the O: $\sqrt{2}\times \sqrt{2}$ unit cell. The system goes through a couple of minima before locating the basin containing the global minimum.
\subsection{\label{sec:level32}$\text{BA}_2\text{MA}\text{Pb}_2\text{Br}_{7}$}
\par
We continue the validation of GO-MHALP on a similar $n=2$ RPP: $\text{BA}_2\text{MA}\text{Pb}_2\text{Br}_{7}$. The obtained similarity measures are summarized in Fig. S10. 
As is the case for $\text{BA}_2\text{Pb}\text{Br}_{4}$, the global minimum of the potential is nearby the experimental structure relaxed with the classical potential with a higher similarity measure compared to the $n=1$ case. A possible rationalization of this fact is that the accuracy of the potential is expected to grow with the number of layers $n$ as the potential is constructed using parameters for a 3D perovskite. 
\par
GO-MHALP correctly predicts that $\text{BA}_2\text{MA}\text{Pb}_2\text{Br}_{7}$ crystallizes in the $\sqrt{2}\times \sqrt{2}$ cell type. 
The low energy region of the PES is not as rich as is the case for $\text{BA}_2\text{Pb}\text{Br}_{4}$ and is surrounded by higher potential barriers. This can be seen in Fig. \ref{fig:ban2_mh}: the first local basin that GO-MHALP found was the one containing the global minimum and higher NPT temperatures were needed to overcome basin barriers. The relative flatness of the PES going from the input to the global minimum is the reason why relaxations of input structures, both by using DFT and classical potentials, resulted in structures near the global minimum with an already large similarity measure. Relaxing the found global minimum with DFT again achieves a slightly better similarity $(0.940)$ compared to relaxing the initial structure with DFT $(0.930)$.
\subsection{\label{sec:level33}$\text{(4AMP)}\text{Pb}\text{Br}_{4}$}
\par
To validate the GO-MHALP workflow on a Q2DP containing a different spacer, we select the Dion-Jacobson type perovskite $\text{(4AMP)}\text{Pb}\text{Br}_{4}$.\cite{mao2018hybrid} 1-(4-Piperidinyl)methanamine ($\text{4AMP}$) structure is shown in Fig. \ref{fig:molecules}. At one molecular end, one nitrogen atom belongs to the aminomethyl unit, while the other is within the piperidinyl ring on the opposite side. In a Dion-Jacobson perovskite, these two nitrogens connect to the inorganic perovskite layers. Therefore, neighbouring organic $\text{4AMP}$ molecules may differently connect to the same layer, i.e. one $\text{4AMP}$ may connect via the aminomethyl unit and its neighbour via the piperidinium unit. To account for this possibility, we considered the following initial configurations:
\begin{itemize}
    \item[(na)] 
    non-alternating; i.e. all neighbouring $\text{4AMP}$ molecules connect to a perovskite layer in the same manner;
    \item[(a1)] 
    all-alternating; i.e. all neighbouring $\text{4AMP}$ molecules connect to a perovskite layer in the opposite manner;
    \item[(a2)] 
    half-alternating; i.e. two neighbouring $\text{4AMP}$ molecules connect to a perovskite layer in the same manner, while the other two connect in the opposite manner.
\end{itemize}
Note that the configuration (a1) is possible for $\sqrt{2}\times\sqrt{2}$ and $2\times 2$ cell types, while (a2) is possible only for $2\times 2$. It is necessary to explicitly include all these configurations in the initial structures since it is extremely unlikely for the molecules to completely reorient during MD. This gives a total number of 12 types of initial structures for GO-MHALP.
\par
A summary of a GO-MHALP run is shown in Fig. \ref{fig:4amp_mh}. After locating ten Q2DP local minima (including the minimum obtained with initial relaxation), GO-MHALP is unable to find a new unique minimum, which results in the MD temperature rising to about 600 K. This high-temperature MD "melts" the inorganic perovskite structure resulting in the newfound local minimum losing its Q2DP character. We deem these "melted" types of structures unphysical predictions as the classical potential is, by construction, well-defined only for Q2DPs and we exclude them from following analysis.
\par
The second point to be noted in Fig. \ref{fig:4amp_mh} is that the energy of the experimentally solved structure relaxed with the classical potential is significantly lower than any of the structures explored by GO-MHALP. The reason for this is that the experimentally resolved structure has more degrees of freedom than the ones describable by configurations in Fig. \ref{fig:4amp_mh}. Specifically, in the notation established in this paper, the experimentally obtained cell is of the $2\sqrt{2}\times\sqrt{2}$ type.\cite{mao2018hybrid} This cell type allows for an intricate pattern of alternating $\text{4AMP}$ orientations: along the shorter cell axis, the manner of connection does not change, while it is altered for every \textit{second} neighbour along the long cell axis.
\par
The summary of GO-MHALP runs for all cell types is shown in Fig. S11. As explained in the last paragraph, all structure types lack the necessary number of degrees of freedom to find the experimental structure relaxed with the classical potential. Relaxing the lowest energy minimum of the $\text{O(a2): }2\times 2$ cell type with DFT actually results in worse similarity $(0.715)$ compared to simply relaxing the initial guess structure with DFT $(0.776)$. However, the energy (as calculated with DFT) of the relaxed $\text{O(a2): }2\times 2$ minimum is significantly lower than the energy of the relaxed initial guess structure ($0.15$ eV per formula unit), meaning that, while it is not close to the experimental structure in a sense, it is a lower energy local minimum. 
\par
To allow GO-MHALP to find the true global minimum, we prepared three input structures with the same cell type as the experimental structure by first removing the one redundant inorganic layer from the $\sqrt{2}\times\sqrt{2}$ cell type, followed by an extension to a $2\sqrt{2}\times\sqrt{2}$ supercell. We considered three $\text{4AMP}$ configurations:
\begin{itemize}
    \item
    nonalternating: all $\text{4AMP}$ molecules connect to the perovskite layer in the same manner;
    \item 
    alternating: the $\text{4AMP}$ molecules alternate the manner of connection to the perovskite layer along the long cell-axis;
    \item 
    experimental-like: the manner of connection of the $\text{4AMP}$ molecules to the perovskite layer is alternated for every \textit{second} neighbour.
\end{itemize}
The results of the GO-MHALP run with these structure types are shown in Fig. \ref{fig:similarity_vs_en_4amp_explike}. The global minimum of the potential (the experimental structure relaxed with the classical potential) is found exclusively for the experimental-like connection pattern. Relaxing the initial guess structure of the experimental cell type with DFT achieves a similarity of $0.74$, while relaxing the found global minimum achieves a remarkable similarity of $0.968$, almost perfectly overlapping the similarity of the experimental structure relaxed with DFT.
\subsection{\label{subsec:level34}$\text{\textit{t}-BA}_2\text{Pb}\text{Br}_{2}\text{I}_{2}$}
We have employed GO-MHALP to predict a previously unknown Q2DP structure and verified its prediction by single-crystal XRD measurements. We used \textit{tert}-butyl ammonium ($\text{\textit{t}-BA}$) as the organic spacer. 
While this is detailed in a separate work \cite{ovcar2022mixed}, we here deepen the discussion of the application of GO-MHALP to that challenging case. The optical measurements indicated that synthesis starting from a one-to-one iodide-bromide stoichiometry results in $\text{\textit{t}-BA}_2\text{Pb}\text{Br}_{2}\text{I}_{2}$, a crystallized $n=1$ RP phase, while syntheses starting from pure bromide or pure iodide stoichiometries do not yield Q2DP structures \cite{ovcar2022mixed}. We confirmed the instability of the pure halide Q2DPs by calculating the formation energies of the global minima found with GO-MHALP as well as by XRD measurements.
\par
To predict the structure of the mixed-halide $\text{\textit{t}-BA}_2\text{Pb}\text{Br}_{2}\text{I}_{2}$, we prepared three types of input structures for GO-MHALP as shown in Fig. \ref{fig:tba_starting_structures} with the corresponding GO-MHALP runs shown in Fig. S12. We see that the minimum corresponding to the experimental structure relaxed with the classical potential is found exclusively starting from the equatorial (axial) bromide (iodine) initial configuration, consistent with the specific halide distribution we found in the structure we resolved experimentally with single-crystal XRD. Relaxing the global minimum with DFT results in a remarkable similarity of 0.967. The relaxed global minimum structure is a local DFT minimum almost isoenergetic to the experimental structure relaxed with DFT (its energy as calculated with DFT is higher by $\approx{4}$ meV/f.u), but it is closer to the experimental structure by $\approx 0.02$ similarity points.

\section{\label{sec:level4}Conclusion}

In this work, we have introduced a workflow for automatic crystal structure prediction of Q2DP structures. To achieve this, we have developed an automatized initial structure guess and classical potential generation and combined them with a variant of the minima hopping algorithm dubbed GO-MHALP. We tested GO-MHALP on well known Q2DP structures: $\text{BA}_2\text{Pb}\text{Br}_4 $, $\text{BA}_2\text{MA}\text{Pb}_2\text{Br}_{7}$ and $\text{(4AMP)}\text{Pb}\text{Br}_{4}$.
We have shown that the global minimum is reliably found by GO-MHALP with a weak dependence on the input structure. 

The case of $\text{(4AMP)}\text{Pb}\text{Br}_{4}$ suggests that it is necessary to start from a structure with the minimum necessary number of degrees of freedom.
We have also shown how GO-MHALP  can be used to predict the structure of a mixed-halide Q2DP: $\text{\textit{t}-BA}_2\text{Pb}\text{Br}_{2}\text{I}_{2}$. 
We confirmed that not only the specific halide distribution was correctly predicted by GO-MHALP, but also that the structural details were predicted very precisely.\cite{ovcar2022mixed}
\par
While the accuracy of the classical potential itself can be significantly improved, within our method it is not necessary for it to be extremely precise. We have shown that the sufficient condition for a very accurate prediction is only that the global minimum of the potential is connected to the DFT global minimum by a DFT local relaxation. However, the potential should be improved in general to provide reliable predictions for any Q2DP.
\par
We also note that our approach provides physically realistic predictions at a low computational price. Assuming one evaluation of energy and forces with a classical potential is $10^6$ times faster than a single SCF calculation, 200 GO-MHALP cycles of 25 ps NPT simulations take about as long as 5 SCF calculations. This is much less than the number of SCF calculations performed in a typical DFT structural relaxation. Therefore, an unbiased structure prediction may be obtained in less time than necessary for two DFT structural relaxations. Since this computational cost is negligible, GO-MHALP may be further improved by using DFT to relax a larger number of predicted structures around the found global minimum.
\par
Compared to previously available minima hopping algorithms, we implemented several improvements: i) the MD temperature is restarted after a structurally unique minimum is found; ii) the replacement scheme, which in combination with i) ensures detailed exploration of local PES basins; iii) employment of an NPT ensemble for the MD part of minima hopping, and iv) inclusion of both cell and atomic coordinate relaxations. We believe that these improvements could be employed in structure prediction problems generally whenever multiple kinds of degrees of freedom (configurational, conformational, combinatorial, etc.) render the exploration of PES particularly difficult, e.g. in soft-matter and molecular crystals.

\section*{\label{sec:level5}Data Availability}
The complete code for generating initial structures, corresponding model potentials and running GO-MHALP is available free of charge at \url{https://github.com/ovcarj/classical-RPP/tree/ase2020}. Using this code all presented data can be regenerated. Derived data are also available from the corresponding author upon reasonable request.
\par See Supplemental Material at [URL will be inserted by publisher] for a benchmark of the accuracy of the classical potentials, a detailed technical description of the structure generation algorithm, calculated radial distribution functions, simulated powder XRD patterns and plots of similarity measures versus potential energies. CIF files of all the initial guess structures and the most relevant found structures are also given in the Supplemental Material.

\section*{\label{sec:level6}Acknowledgments}
This work was supported by the PZS-2019-02-2068 project financed by the “Research Cooperability” Program of the Croatian Science Foundation and European Union from the European Social Fund under the Operational Programme Efficient Human Resources 2014-2020 and NSFC project 6207032617.

\bibliography{input}

\begin{thebibliography}{70}%
\makeatletter
\providecommand \@ifxundefined [1]{%
 \@ifx{#1\undefined}
}%
\providecommand \@ifnum [1]{%
 \ifnum #1\expandafter \@firstoftwo
 \else \expandafter \@secondoftwo
 \fi
}%
\providecommand \@ifx [1]{%
 \ifx #1\expandafter \@firstoftwo
 \else \expandafter \@secondoftwo
 \fi
}%
\providecommand \natexlab [1]{#1}%
\providecommand \enquote  [1]{``#1''}%
\providecommand \bibnamefont  [1]{#1}%
\providecommand \bibfnamefont [1]{#1}%
\providecommand \citenamefont [1]{#1}%
\providecommand \href@noop [0]{\@secondoftwo}%
\providecommand \href [0]{\begingroup \@sanitize@url \@href}%
\providecommand \@href[1]{\@@startlink{#1}\@@href}%
\providecommand \@@href[1]{\endgroup#1\@@endlink}%
\providecommand \@sanitize@url [0]{\catcode `\\12\catcode `\$12\catcode
  `\&12\catcode `\#12\catcode `\^12\catcode `\_12\catcode `\%12\relax}%
\providecommand \@@startlink[1]{}%
\providecommand \@@endlink[0]{}%
\providecommand \url  [0]{\begingroup\@sanitize@url \@url }%
\providecommand \@url [1]{\endgroup\@href {#1}{\urlprefix }}%
\providecommand \urlprefix  [0]{URL }%
\providecommand \Eprint [0]{\href }%
\providecommand \doibase [0]{https://doi.org/}%
\providecommand \selectlanguage [0]{\@gobble}%
\providecommand \bibinfo  [0]{\@secondoftwo}%
\providecommand \bibfield  [0]{\@secondoftwo}%
\providecommand \translation [1]{[#1]}%
\providecommand \BibitemOpen [0]{}%
\providecommand \bibitemStop [0]{}%
\providecommand \bibitemNoStop [0]{.\EOS\space}%
\providecommand \EOS [0]{\spacefactor3000\relax}%
\providecommand \BibitemShut  [1]{\csname bibitem#1\endcsname}%
\let\auto@bib@innerbib\@empty
\bibitem [{\citenamefont {Jena}\ \emph {et~al.}(2019)\citenamefont {Jena},
  \citenamefont {Kulkarni},\ and\ \citenamefont {Miyasaka}}]{jena2019halide}%
  \BibitemOpen
  \bibfield  {author} {\bibinfo {author} {\bibfnamefont {A.~K.}\ \bibnamefont
  {Jena}}, \bibinfo {author} {\bibfnamefont {A.}~\bibnamefont {Kulkarni}},\
  and\ \bibinfo {author} {\bibfnamefont {T.}~\bibnamefont {Miyasaka}},\
  }\bibfield  {title} {\bibinfo {title} {Halide perovskite photovoltaics:
  background, status, and future prospects},\ }\href@noop {} {\bibfield
  {journal} {\bibinfo  {journal} {Chemical reviews}\ }\textbf {\bibinfo
  {volume} {119}},\ \bibinfo {pages} {3036} (\bibinfo {year}
  {2019})}\BibitemShut {NoStop}%
\bibitem [{\citenamefont {Lu}\ \emph {et~al.}(2019)\citenamefont {Lu},
  \citenamefont {Zhang}, \citenamefont {Wang}, \citenamefont {Guo},
  \citenamefont {Yu},\ and\ \citenamefont {Rogach}}]{lu2019metal}%
  \BibitemOpen
  \bibfield  {author} {\bibinfo {author} {\bibfnamefont {M.}~\bibnamefont
  {Lu}}, \bibinfo {author} {\bibfnamefont {Y.}~\bibnamefont {Zhang}}, \bibinfo
  {author} {\bibfnamefont {S.}~\bibnamefont {Wang}}, \bibinfo {author}
  {\bibfnamefont {J.}~\bibnamefont {Guo}}, \bibinfo {author} {\bibfnamefont
  {W.~W.}\ \bibnamefont {Yu}},\ and\ \bibinfo {author} {\bibfnamefont {A.~L.}\
  \bibnamefont {Rogach}},\ }\bibfield  {title} {\bibinfo {title} {Metal halide
  perovskite light-emitting devices: promising technology for next-generation
  displays},\ }\href@noop {} {\bibfield  {journal} {\bibinfo  {journal}
  {Advanced Functional Materials}\ }\textbf {\bibinfo {volume} {29}},\ \bibinfo
  {pages} {1902008} (\bibinfo {year} {2019})}\BibitemShut {NoStop}%
\bibitem [{\citenamefont {Mao}\ \emph {et~al.}(2018{\natexlab{a}})\citenamefont
  {Mao}, \citenamefont {Stoumpos},\ and\ \citenamefont
  {Kanatzidis}}]{mao2018two}%
  \BibitemOpen
  \bibfield  {author} {\bibinfo {author} {\bibfnamefont {L.}~\bibnamefont
  {Mao}}, \bibinfo {author} {\bibfnamefont {C.~C.}\ \bibnamefont {Stoumpos}},\
  and\ \bibinfo {author} {\bibfnamefont {M.~G.}\ \bibnamefont {Kanatzidis}},\
  }\bibfield  {title} {\bibinfo {title} {Two-dimensional hybrid halide
  perovskites: principles and promises},\ }\href@noop {} {\bibfield  {journal}
  {\bibinfo  {journal} {Journal of the American Chemical Society}\ }\textbf
  {\bibinfo {volume} {141}},\ \bibinfo {pages} {1171} (\bibinfo {year}
  {2018}{\natexlab{a}})}\BibitemShut {NoStop}%
\bibitem [{\citenamefont {Chen}\ \emph {et~al.}(2019)\citenamefont {Chen},
  \citenamefont {Guo}, \citenamefont {Wertz},\ and\ \citenamefont
  {Shi}}]{chen2019merits}%
  \BibitemOpen
  \bibfield  {author} {\bibinfo {author} {\bibfnamefont {Z.}~\bibnamefont
  {Chen}}, \bibinfo {author} {\bibfnamefont {Y.}~\bibnamefont {Guo}}, \bibinfo
  {author} {\bibfnamefont {E.}~\bibnamefont {Wertz}},\ and\ \bibinfo {author}
  {\bibfnamefont {J.}~\bibnamefont {Shi}},\ }\bibfield  {title} {\bibinfo
  {title} {Merits and challenges of ruddlesden--popper soft halide perovskites
  in electro-optics and optoelectronics},\ }\href@noop {} {\bibfield  {journal}
  {\bibinfo  {journal} {Advanced Materials}\ }\textbf {\bibinfo {volume}
  {31}},\ \bibinfo {pages} {1803514} (\bibinfo {year} {2019})}\BibitemShut
  {NoStop}%
\bibitem [{\citenamefont {Ahmad}\ \emph {et~al.}(2019)\citenamefont {Ahmad},
  \citenamefont {Fu}, \citenamefont {Yu}, \citenamefont {Yang}, \citenamefont
  {Liu}, \citenamefont {Wang}, \citenamefont {Wang}, \citenamefont {Guo},\ and\
  \citenamefont {Li}}]{ahmad2019dion}%
  \BibitemOpen
  \bibfield  {author} {\bibinfo {author} {\bibfnamefont {S.}~\bibnamefont
  {Ahmad}}, \bibinfo {author} {\bibfnamefont {P.}~\bibnamefont {Fu}}, \bibinfo
  {author} {\bibfnamefont {S.}~\bibnamefont {Yu}}, \bibinfo {author}
  {\bibfnamefont {Q.}~\bibnamefont {Yang}}, \bibinfo {author} {\bibfnamefont
  {X.}~\bibnamefont {Liu}}, \bibinfo {author} {\bibfnamefont {X.}~\bibnamefont
  {Wang}}, \bibinfo {author} {\bibfnamefont {X.}~\bibnamefont {Wang}}, \bibinfo
  {author} {\bibfnamefont {X.}~\bibnamefont {Guo}},\ and\ \bibinfo {author}
  {\bibfnamefont {C.}~\bibnamefont {Li}},\ }\bibfield  {title} {\bibinfo
  {title} {Dion-jacobson phase 2d layered perovskites for solar cells with
  ultrahigh stability},\ }\href@noop {} {\bibfield  {journal} {\bibinfo
  {journal} {Joule}\ }\textbf {\bibinfo {volume} {3}},\ \bibinfo {pages} {794}
  (\bibinfo {year} {2019})}\BibitemShut {NoStop}%
\bibitem [{\citenamefont {David}\ \emph {et~al.}(2002)\citenamefont {David},
  \citenamefont {Shankland} \emph {et~al.}}]{david2002structure}%
  \BibitemOpen
  \bibfield  {author} {\bibinfo {author} {\bibfnamefont {W.~I.}\ \bibnamefont
  {David}}, \bibinfo {author} {\bibfnamefont {K.}~\bibnamefont {Shankland}},
  \emph {et~al.},\ }\href@noop {} {\emph {\bibinfo {title} {Structure
  determination from powder diffraction data}}},\ Vol.~\bibinfo {volume} {13}\
  (\bibinfo  {publisher} {Oxford University Press on Demand},\ \bibinfo {year}
  {2002})\BibitemShut {NoStop}%
\bibitem [{\citenamefont {Favre-Nicolin}\ and\ \citenamefont
  {{\v{C}}ern{\`y}}(2002)}]{favre2002fox}%
  \BibitemOpen
  \bibfield  {author} {\bibinfo {author} {\bibfnamefont {V.}~\bibnamefont
  {Favre-Nicolin}}\ and\ \bibinfo {author} {\bibfnamefont {R.}~\bibnamefont
  {{\v{C}}ern{\`y}}},\ }\bibfield  {title} {\bibinfo {title} {Fox,free objects
  for crystallography': a modular approach to ab initio structure determination
  from powder diffraction},\ }\href@noop {} {\bibfield  {journal} {\bibinfo
  {journal} {Journal of Applied Crystallography}\ }\textbf {\bibinfo {volume}
  {35}},\ \bibinfo {pages} {734} (\bibinfo {year} {2002})}\BibitemShut
  {NoStop}%
\bibitem [{\citenamefont {David}\ \emph {et~al.}(2006)\citenamefont {David},
  \citenamefont {Shankland}, \citenamefont {Van De~Streek}, \citenamefont
  {Pidcock}, \citenamefont {Motherwell},\ and\ \citenamefont
  {Cole}}]{david2006dash}%
  \BibitemOpen
  \bibfield  {author} {\bibinfo {author} {\bibfnamefont {W.~I.}\ \bibnamefont
  {David}}, \bibinfo {author} {\bibfnamefont {K.}~\bibnamefont {Shankland}},
  \bibinfo {author} {\bibfnamefont {J.}~\bibnamefont {Van De~Streek}}, \bibinfo
  {author} {\bibfnamefont {E.}~\bibnamefont {Pidcock}}, \bibinfo {author}
  {\bibfnamefont {W.~S.}\ \bibnamefont {Motherwell}},\ and\ \bibinfo {author}
  {\bibfnamefont {J.~C.}\ \bibnamefont {Cole}},\ }\bibfield  {title} {\bibinfo
  {title} {Dash: a program for crystal structure determination from powder
  diffraction data},\ }\href@noop {} {\bibfield  {journal} {\bibinfo  {journal}
  {Journal of applied crystallography}\ }\textbf {\bibinfo {volume} {39}},\
  \bibinfo {pages} {910} (\bibinfo {year} {2006})}\BibitemShut {NoStop}%
\bibitem [{\citenamefont {Altomare}\ \emph {et~al.}(2013)\citenamefont
  {Altomare}, \citenamefont {Cuocci}, \citenamefont {Giacovazzo}, \citenamefont
  {Moliterni}, \citenamefont {Rizzi}, \citenamefont {Corriero},\ and\
  \citenamefont {Falcicchio}}]{altomare2013expo2013}%
  \BibitemOpen
  \bibfield  {author} {\bibinfo {author} {\bibfnamefont {A.}~\bibnamefont
  {Altomare}}, \bibinfo {author} {\bibfnamefont {C.}~\bibnamefont {Cuocci}},
  \bibinfo {author} {\bibfnamefont {C.}~\bibnamefont {Giacovazzo}}, \bibinfo
  {author} {\bibfnamefont {A.}~\bibnamefont {Moliterni}}, \bibinfo {author}
  {\bibfnamefont {R.}~\bibnamefont {Rizzi}}, \bibinfo {author} {\bibfnamefont
  {N.}~\bibnamefont {Corriero}},\ and\ \bibinfo {author} {\bibfnamefont
  {A.}~\bibnamefont {Falcicchio}},\ }\bibfield  {title} {\bibinfo {title}
  {Expo2013: a kit of tools for phasing crystal structures from powder data},\
  }\href@noop {} {\bibfield  {journal} {\bibinfo  {journal} {Journal of Applied
  Crystallography}\ }\textbf {\bibinfo {volume} {46}},\ \bibinfo {pages} {1231}
  (\bibinfo {year} {2013})}\BibitemShut {NoStop}%
\bibitem [{\citenamefont {C{\v{C}}ern{\`y}}(2017)}]{cvcerny2017crystal}%
  \BibitemOpen
  \bibfield  {author} {\bibinfo {author} {\bibfnamefont {R.}~\bibnamefont
  {C{\v{C}}ern{\`y}}},\ }\bibfield  {title} {\bibinfo {title} {Crystal
  structures from powder diffraction: Principles, difficulties and progress.},\
  }\href@noop {} {\bibfield  {journal} {\bibinfo  {journal} {Crystals
  (2073-4352)}\ }\textbf {\bibinfo {volume} {7}} (\bibinfo {year}
  {2017})}\BibitemShut {NoStop}%
\bibitem [{\citenamefont {Jones}\ \emph {et~al.}(2017)\citenamefont {Jones},
  \citenamefont {R{\"o}thel}, \citenamefont {Lassnig}, \citenamefont
  {Bedoya-Mart{\'\i}nez}, \citenamefont {Christian}, \citenamefont {Salzmann},
  \citenamefont {Kunert}, \citenamefont {Winkler},\ and\ \citenamefont
  {Resel}}]{jones2017solution}%
  \BibitemOpen
  \bibfield  {author} {\bibinfo {author} {\bibfnamefont {A.~O.}\ \bibnamefont
  {Jones}}, \bibinfo {author} {\bibfnamefont {C.}~\bibnamefont {R{\"o}thel}},
  \bibinfo {author} {\bibfnamefont {R.}~\bibnamefont {Lassnig}}, \bibinfo
  {author} {\bibfnamefont {O.}~\bibnamefont {Bedoya-Mart{\'\i}nez}}, \bibinfo
  {author} {\bibfnamefont {P.}~\bibnamefont {Christian}}, \bibinfo {author}
  {\bibfnamefont {I.}~\bibnamefont {Salzmann}}, \bibinfo {author}
  {\bibfnamefont {B.}~\bibnamefont {Kunert}}, \bibinfo {author} {\bibfnamefont
  {A.}~\bibnamefont {Winkler}},\ and\ \bibinfo {author} {\bibfnamefont
  {R.}~\bibnamefont {Resel}},\ }\bibfield  {title} {\bibinfo {title} {Solution
  of an elusive pigment crystal structure from a thin film: a combined x-ray
  diffraction and computational study},\ }\href@noop {} {\bibfield  {journal}
  {\bibinfo  {journal} {CrystEngComm}\ }\textbf {\bibinfo {volume} {19}},\
  \bibinfo {pages} {1902} (\bibinfo {year} {2017})}\BibitemShut {NoStop}%
\bibitem [{\citenamefont {Jones}\ \emph {et~al.}(2016)\citenamefont {Jones},
  \citenamefont {Chattopadhyay}, \citenamefont {Geerts},\ and\ \citenamefont
  {Resel}}]{jones2016substrate}%
  \BibitemOpen
  \bibfield  {author} {\bibinfo {author} {\bibfnamefont {A.~O.}\ \bibnamefont
  {Jones}}, \bibinfo {author} {\bibfnamefont {B.}~\bibnamefont
  {Chattopadhyay}}, \bibinfo {author} {\bibfnamefont {Y.~H.}\ \bibnamefont
  {Geerts}},\ and\ \bibinfo {author} {\bibfnamefont {R.}~\bibnamefont
  {Resel}},\ }\bibfield  {title} {\bibinfo {title} {Substrate-induced and
  thin-film phases: Polymorphism of organic materials on surfaces},\
  }\href@noop {} {\bibfield  {journal} {\bibinfo  {journal} {Advanced
  functional materials}\ }\textbf {\bibinfo {volume} {26}},\ \bibinfo {pages}
  {2233} (\bibinfo {year} {2016})}\BibitemShut {NoStop}%
\bibitem [{\citenamefont {Tao}\ \emph {et~al.}(2021)\citenamefont {Tao},
  \citenamefont {Xu}, \citenamefont {Li},\ and\ \citenamefont
  {Lu}}]{tao2021machine}%
  \BibitemOpen
  \bibfield  {author} {\bibinfo {author} {\bibfnamefont {Q.}~\bibnamefont
  {Tao}}, \bibinfo {author} {\bibfnamefont {P.}~\bibnamefont {Xu}}, \bibinfo
  {author} {\bibfnamefont {M.}~\bibnamefont {Li}},\ and\ \bibinfo {author}
  {\bibfnamefont {W.}~\bibnamefont {Lu}},\ }\bibfield  {title} {\bibinfo
  {title} {Machine learning for perovskite materials design and discovery},\
  }\href@noop {} {\bibfield  {journal} {\bibinfo  {journal} {npj Computational
  Materials}\ }\textbf {\bibinfo {volume} {7}},\ \bibinfo {pages} {1} (\bibinfo
  {year} {2021})}\BibitemShut {NoStop}%
\bibitem [{\citenamefont {Li}\ \emph {et~al.}(2021{\natexlab{a}})\citenamefont
  {Li}, \citenamefont {Tao}, \citenamefont {Xu}, \citenamefont {Yang},
  \citenamefont {Lu},\ and\ \citenamefont {Li}}]{li2021studies}%
  \BibitemOpen
  \bibfield  {author} {\bibinfo {author} {\bibfnamefont {L.}~\bibnamefont
  {Li}}, \bibinfo {author} {\bibfnamefont {Q.}~\bibnamefont {Tao}}, \bibinfo
  {author} {\bibfnamefont {P.}~\bibnamefont {Xu}}, \bibinfo {author}
  {\bibfnamefont {X.}~\bibnamefont {Yang}}, \bibinfo {author} {\bibfnamefont
  {W.}~\bibnamefont {Lu}},\ and\ \bibinfo {author} {\bibfnamefont
  {M.}~\bibnamefont {Li}},\ }\bibfield  {title} {\bibinfo {title} {Studies on
  the regularity of perovskite formation via machine learning},\ }\href@noop {}
  {\bibfield  {journal} {\bibinfo  {journal} {Computational Materials Science}\
  }\textbf {\bibinfo {volume} {199}},\ \bibinfo {pages} {110712} (\bibinfo
  {year} {2021}{\natexlab{a}})}\BibitemShut {NoStop}%
\bibitem [{\citenamefont {G{\'o}mez-Peralta}\ and\ \citenamefont
  {Bokhimi}(2021)}]{gomez2021ternary}%
  \BibitemOpen
  \bibfield  {author} {\bibinfo {author} {\bibfnamefont {J.}~\bibnamefont
  {G{\'o}mez-Peralta}}\ and\ \bibinfo {author} {\bibfnamefont {X.}~\bibnamefont
  {Bokhimi}},\ }\bibfield  {title} {\bibinfo {title} {Ternary halide
  perovskites for possible optoelectronic applications revealed by artificial
  intelligence and dft calculations},\ }\href@noop {} {\bibfield  {journal}
  {\bibinfo  {journal} {Materials Chemistry and Physics}\ }\textbf {\bibinfo
  {volume} {267}},\ \bibinfo {pages} {124710} (\bibinfo {year}
  {2021})}\BibitemShut {NoStop}%
\bibitem [{\citenamefont {Jahanbakhshi}\ \emph {et~al.}(2021)\citenamefont
  {Jahanbakhshi}, \citenamefont {Mladenovi{\'c}}, \citenamefont {Dankl},
  \citenamefont {Boziki}, \citenamefont {Ahlawat},\ and\ \citenamefont
  {Rothlisberger}}]{jahanbakhshi2021organic}%
  \BibitemOpen
  \bibfield  {author} {\bibinfo {author} {\bibfnamefont {F.}~\bibnamefont
  {Jahanbakhshi}}, \bibinfo {author} {\bibfnamefont {M.}~\bibnamefont
  {Mladenovi{\'c}}}, \bibinfo {author} {\bibfnamefont {M.}~\bibnamefont
  {Dankl}}, \bibinfo {author} {\bibfnamefont {A.}~\bibnamefont {Boziki}},
  \bibinfo {author} {\bibfnamefont {P.}~\bibnamefont {Ahlawat}},\ and\ \bibinfo
  {author} {\bibfnamefont {U.}~\bibnamefont {Rothlisberger}},\ }\bibfield
  {title} {\bibinfo {title} {Organic spacers in 2d perovskites: General trends
  and structure-property relationships from computational studies},\
  }\href@noop {} {\bibfield  {journal} {\bibinfo  {journal} {Helvetica Chimica
  Acta}\ }\textbf {\bibinfo {volume} {104}},\ \bibinfo {pages} {e2000232}
  (\bibinfo {year} {2021})}\BibitemShut {NoStop}%
\bibitem [{\citenamefont {Lyu}\ \emph {et~al.}(2021)\citenamefont {Lyu},
  \citenamefont {Moore}, \citenamefont {Liu}, \citenamefont {Yu},\ and\
  \citenamefont {Wu}}]{lyu2021predictive}%
  \BibitemOpen
  \bibfield  {author} {\bibinfo {author} {\bibfnamefont {R.}~\bibnamefont
  {Lyu}}, \bibinfo {author} {\bibfnamefont {C.~E.}\ \bibnamefont {Moore}},
  \bibinfo {author} {\bibfnamefont {T.}~\bibnamefont {Liu}}, \bibinfo {author}
  {\bibfnamefont {Y.}~\bibnamefont {Yu}},\ and\ \bibinfo {author}
  {\bibfnamefont {Y.}~\bibnamefont {Wu}},\ }\bibfield  {title} {\bibinfo
  {title} {Predictive design model for low-dimensional organic--inorganic
  halide perovskites assisted by machine learning},\ }\href@noop {} {\bibfield
  {journal} {\bibinfo  {journal} {Journal of the American Chemical Society}\
  }\textbf {\bibinfo {volume} {143}},\ \bibinfo {pages} {12766} (\bibinfo
  {year} {2021})}\BibitemShut {NoStop}%
\bibitem [{\citenamefont {Price}\ \emph {et~al.}(2020)\citenamefont {Price},
  \citenamefont {Blancon}, \citenamefont {Mohite},\ and\ \citenamefont
  {Shenoy}}]{price2020interfacial}%
  \BibitemOpen
  \bibfield  {author} {\bibinfo {author} {\bibfnamefont {C.~C.}\ \bibnamefont
  {Price}}, \bibinfo {author} {\bibfnamefont {J.-C.}\ \bibnamefont {Blancon}},
  \bibinfo {author} {\bibfnamefont {A.~D.}\ \bibnamefont {Mohite}},\ and\
  \bibinfo {author} {\bibfnamefont {V.~B.}\ \bibnamefont {Shenoy}},\ }\bibfield
   {title} {\bibinfo {title} {Interfacial electromechanics predicts phase
  behavior of 2d hybrid halide perovskites},\ }\href@noop {} {\bibfield
  {journal} {\bibinfo  {journal} {ACS nano}\ }\textbf {\bibinfo {volume}
  {14}},\ \bibinfo {pages} {3353} (\bibinfo {year} {2020})}\BibitemShut
  {NoStop}%
\bibitem [{\citenamefont {Goedecker}(2004)}]{goedecker2004minima}%
  \BibitemOpen
  \bibfield  {author} {\bibinfo {author} {\bibfnamefont {S.}~\bibnamefont
  {Goedecker}},\ }\bibfield  {title} {\bibinfo {title} {Minima hopping: An
  efficient search method for the global minimum of the potential energy
  surface of complex molecular systems},\ }\href@noop {} {\bibfield  {journal}
  {\bibinfo  {journal} {The Journal of chemical physics}\ }\textbf {\bibinfo
  {volume} {120}},\ \bibinfo {pages} {9911} (\bibinfo {year}
  {2004})}\BibitemShut {NoStop}%
\bibitem [{\citenamefont {Amsler}\ and\ \citenamefont
  {Goedecker}(2010)}]{amsler2010crystal}%
  \BibitemOpen
  \bibfield  {author} {\bibinfo {author} {\bibfnamefont {M.}~\bibnamefont
  {Amsler}}\ and\ \bibinfo {author} {\bibfnamefont {S.}~\bibnamefont
  {Goedecker}},\ }\bibfield  {title} {\bibinfo {title} {Crystal structure
  prediction using the minima hopping method},\ }\href@noop {} {\bibfield
  {journal} {\bibinfo  {journal} {The Journal of chemical physics}\ }\textbf
  {\bibinfo {volume} {133}},\ \bibinfo {pages} {224104} (\bibinfo {year}
  {2010})}\BibitemShut {NoStop}%
\bibitem [{\citenamefont {Peterson}(2014)}]{peterson2014global}%
  \BibitemOpen
  \bibfield  {author} {\bibinfo {author} {\bibfnamefont {A.~A.}\ \bibnamefont
  {Peterson}},\ }\bibfield  {title} {\bibinfo {title} {Global optimization of
  adsorbate--surface structures while preserving molecular identity},\
  }\href@noop {} {\bibfield  {journal} {\bibinfo  {journal} {Topics in
  Catalysis}\ }\textbf {\bibinfo {volume} {57}},\ \bibinfo {pages} {40}
  (\bibinfo {year} {2014})}\BibitemShut {NoStop}%
\bibitem [{\citenamefont {Wang}\ \emph {et~al.}(2012)\citenamefont {Wang},
  \citenamefont {Lv}, \citenamefont {Zhu},\ and\ \citenamefont
  {Ma}}]{wang2012calypso}%
  \BibitemOpen
  \bibfield  {author} {\bibinfo {author} {\bibfnamefont {Y.}~\bibnamefont
  {Wang}}, \bibinfo {author} {\bibfnamefont {J.}~\bibnamefont {Lv}}, \bibinfo
  {author} {\bibfnamefont {L.}~\bibnamefont {Zhu}},\ and\ \bibinfo {author}
  {\bibfnamefont {Y.}~\bibnamefont {Ma}},\ }\bibfield  {title} {\bibinfo
  {title} {Calypso: A method for crystal structure prediction},\ }\href@noop {}
  {\bibfield  {journal} {\bibinfo  {journal} {Computer Physics Communications}\
  }\textbf {\bibinfo {volume} {183}},\ \bibinfo {pages} {2063} (\bibinfo {year}
  {2012})}\BibitemShut {NoStop}%
\bibitem [{\citenamefont {Glass}\ \emph {et~al.}(2006)\citenamefont {Glass},
  \citenamefont {Oganov},\ and\ \citenamefont {Hansen}}]{glass2006uspex}%
  \BibitemOpen
  \bibfield  {author} {\bibinfo {author} {\bibfnamefont {C.~W.}\ \bibnamefont
  {Glass}}, \bibinfo {author} {\bibfnamefont {A.~R.}\ \bibnamefont {Oganov}},\
  and\ \bibinfo {author} {\bibfnamefont {N.}~\bibnamefont {Hansen}},\
  }\bibfield  {title} {\bibinfo {title} {Uspex—evolutionary crystal structure
  prediction},\ }\href@noop {} {\bibfield  {journal} {\bibinfo  {journal}
  {Computer physics communications}\ }\textbf {\bibinfo {volume} {175}},\
  \bibinfo {pages} {713} (\bibinfo {year} {2006})}\BibitemShut {NoStop}%
\bibitem [{\citenamefont {Ov\v{c}ar}\ \emph {et~al.}(2022)\citenamefont
  {Ov\v{c}ar}, \citenamefont {Leung}, \citenamefont {Grisanti}, \citenamefont
  {Skoko}, \citenamefont {Vranki{\'c}}, \citenamefont {Low}, \citenamefont
  {Wang}, \citenamefont {You}, \citenamefont {Ahn}, \citenamefont
  {Lon\v{c}ari{\'c}}, \citenamefont {Djuri\v{s}i{\'c}},\ and\ \citenamefont
  {Popovi{\'c}}}]{ovcar2022mixed}%
  \BibitemOpen
  \bibfield  {author} {\bibinfo {author} {\bibfnamefont {J.}~\bibnamefont
  {Ov\v{c}ar}}, \bibinfo {author} {\bibfnamefont {T.}~\bibnamefont {Leung}},
  \bibinfo {author} {\bibfnamefont {L.}~\bibnamefont {Grisanti}}, \bibinfo
  {author} {\bibfnamefont {v.}~\bibnamefont {Skoko}}, \bibinfo {author}
  {\bibfnamefont {M.}~\bibnamefont {Vranki{\'c}}}, \bibinfo {author}
  {\bibfnamefont {K.-H.}\ \bibnamefont {Low}}, \bibinfo {author} {\bibfnamefont
  {S.}~\bibnamefont {Wang}}, \bibinfo {author} {\bibfnamefont {P.-Y.}\
  \bibnamefont {You}}, \bibinfo {author} {\bibfnamefont {H.}~\bibnamefont
  {Ahn}}, \bibinfo {author} {\bibfnamefont {I.}~\bibnamefont
  {Lon\v{c}ari{\'c}}}, \bibinfo {author} {\bibfnamefont {A.}~\bibnamefont
  {Djuri\v{s}i{\'c}}},\ and\ \bibinfo {author} {\bibfnamefont {J.}~\bibnamefont
  {Popovi{\'c}}},\ }\bibfield  {title} {\bibinfo {title} {Mixed halide ordering
  as a tool for the stabilization of ruddlesden-popper structures},\
  }\href@noop {} {\bibfield  {journal} {\bibinfo  {journal} {Chemistry of
  Materials}\ } (\bibinfo {year} {2022})}\BibitemShut {NoStop}%
\bibitem [{\citenamefont {Li}\ \emph {et~al.}(2021{\natexlab{b}})\citenamefont
  {Li}, \citenamefont {Hoffman},\ and\ \citenamefont {Kanatzidis}}]{li20212d}%
  \BibitemOpen
  \bibfield  {author} {\bibinfo {author} {\bibfnamefont {X.}~\bibnamefont
  {Li}}, \bibinfo {author} {\bibfnamefont {J.~M.}\ \bibnamefont {Hoffman}},\
  and\ \bibinfo {author} {\bibfnamefont {M.~G.}\ \bibnamefont {Kanatzidis}},\
  }\bibfield  {title} {\bibinfo {title} {The 2d halide perovskite rulebook: How
  the spacer influences everything from the structure to optoelectronic device
  efficiency},\ }\href@noop {} {\bibfield  {journal} {\bibinfo  {journal}
  {Chemical Reviews}\ }\textbf {\bibinfo {volume} {121}},\ \bibinfo {pages}
  {2230} (\bibinfo {year} {2021}{\natexlab{b}})}\BibitemShut {NoStop}%
\bibitem [{\citenamefont {Matsui}\ \emph {et~al.}(1987)\citenamefont {Matsui},
  \citenamefont {Akaogi},\ and\ \citenamefont
  {Matsumoto}}]{matsui1987computational}%
  \BibitemOpen
  \bibfield  {author} {\bibinfo {author} {\bibfnamefont {M.}~\bibnamefont
  {Matsui}}, \bibinfo {author} {\bibfnamefont {M.}~\bibnamefont {Akaogi}},\
  and\ \bibinfo {author} {\bibfnamefont {T.}~\bibnamefont {Matsumoto}},\
  }\bibfield  {title} {\bibinfo {title} {Computational model of the structural
  and elastic properties of the ilmenite and perovskite phases of mgsio 3},\
  }\href@noop {} {\bibfield  {journal} {\bibinfo  {journal} {Physics and
  Chemistry of Minerals}\ }\textbf {\bibinfo {volume} {14}},\ \bibinfo {pages}
  {101} (\bibinfo {year} {1987})}\BibitemShut {NoStop}%
\bibitem [{\citenamefont {Saba}\ and\ \citenamefont
  {Mattoni}(2014)}]{saba2014effect}%
  \BibitemOpen
  \bibfield  {author} {\bibinfo {author} {\bibfnamefont {M.~I.}\ \bibnamefont
  {Saba}}\ and\ \bibinfo {author} {\bibfnamefont {A.}~\bibnamefont {Mattoni}},\
  }\bibfield  {title} {\bibinfo {title} {Effect of thermodynamics and curvature
  on the crystallinity of p3ht thin films on zno: insights from atomistic
  simulations},\ }\href@noop {} {\bibfield  {journal} {\bibinfo  {journal} {The
  Journal of Physical Chemistry C}\ }\textbf {\bibinfo {volume} {118}},\
  \bibinfo {pages} {4687} (\bibinfo {year} {2014})}\BibitemShut {NoStop}%
\bibitem [{\citenamefont {Mattoni}\ \emph {et~al.}(2015)\citenamefont
  {Mattoni}, \citenamefont {Filippetti}, \citenamefont {Saba},\ and\
  \citenamefont {Delugas}}]{mattoni2015methylammonium}%
  \BibitemOpen
  \bibfield  {author} {\bibinfo {author} {\bibfnamefont {A.}~\bibnamefont
  {Mattoni}}, \bibinfo {author} {\bibfnamefont {A.}~\bibnamefont {Filippetti}},
  \bibinfo {author} {\bibfnamefont {M.}~\bibnamefont {Saba}},\ and\ \bibinfo
  {author} {\bibfnamefont {P.}~\bibnamefont {Delugas}},\ }\bibfield  {title}
  {\bibinfo {title} {Methylammonium rotational dynamics in lead halide
  perovskite by classical molecular dynamics: the role of temperature},\
  }\href@noop {} {\bibfield  {journal} {\bibinfo  {journal} {The Journal of
  Physical Chemistry C}\ }\textbf {\bibinfo {volume} {119}},\ \bibinfo {pages}
  {17421} (\bibinfo {year} {2015})}\BibitemShut {NoStop}%
\bibitem [{\citenamefont {Hata}\ \emph {et~al.}(2017)\citenamefont {Hata},
  \citenamefont {Giorgi}, \citenamefont {Yamashita}, \citenamefont {Caddeo},\
  and\ \citenamefont {Mattoni}}]{hata2017development}%
  \BibitemOpen
  \bibfield  {author} {\bibinfo {author} {\bibfnamefont {T.}~\bibnamefont
  {Hata}}, \bibinfo {author} {\bibfnamefont {G.}~\bibnamefont {Giorgi}},
  \bibinfo {author} {\bibfnamefont {K.}~\bibnamefont {Yamashita}}, \bibinfo
  {author} {\bibfnamefont {C.}~\bibnamefont {Caddeo}},\ and\ \bibinfo {author}
  {\bibfnamefont {A.}~\bibnamefont {Mattoni}},\ }\bibfield  {title} {\bibinfo
  {title} {Development of a classical interatomic potential for mapbbr3},\
  }\href@noop {} {\bibfield  {journal} {\bibinfo  {journal} {The Journal of
  Physical Chemistry C}\ }\textbf {\bibinfo {volume} {121}},\ \bibinfo {pages}
  {3724} (\bibinfo {year} {2017})}\BibitemShut {NoStop}%
\bibitem [{\citenamefont {Fridriksson}\ \emph {et~al.}(2020)\citenamefont
  {Fridriksson}, \citenamefont {Van Der~Meer}, \citenamefont {De~Haas},\ and\
  \citenamefont {Grozema}}]{fridriksson2020tuning}%
  \BibitemOpen
  \bibfield  {author} {\bibinfo {author} {\bibfnamefont {M.~B.}\ \bibnamefont
  {Fridriksson}}, \bibinfo {author} {\bibfnamefont {N.}~\bibnamefont {Van
  Der~Meer}}, \bibinfo {author} {\bibfnamefont {J.}~\bibnamefont {De~Haas}},\
  and\ \bibinfo {author} {\bibfnamefont {F.~C.}\ \bibnamefont {Grozema}},\
  }\bibfield  {title} {\bibinfo {title} {Tuning the structural rigidity of
  two-dimensional ruddlesden--popper perovskites through the organic cation},\
  }\href@noop {} {\bibfield  {journal} {\bibinfo  {journal} {The Journal of
  Physical Chemistry C}\ }\textbf {\bibinfo {volume} {124}},\ \bibinfo {pages}
  {28201} (\bibinfo {year} {2020})}\BibitemShut {NoStop}%
\bibitem [{\citenamefont {Wang}\ \emph {et~al.}(2004)\citenamefont {Wang},
  \citenamefont {Wolf}, \citenamefont {Caldwell}, \citenamefont {Kollman},\
  and\ \citenamefont {Case}}]{wang2004development}%
  \BibitemOpen
  \bibfield  {author} {\bibinfo {author} {\bibfnamefont {J.}~\bibnamefont
  {Wang}}, \bibinfo {author} {\bibfnamefont {R.~M.}\ \bibnamefont {Wolf}},
  \bibinfo {author} {\bibfnamefont {J.~W.}\ \bibnamefont {Caldwell}}, \bibinfo
  {author} {\bibfnamefont {P.~A.}\ \bibnamefont {Kollman}},\ and\ \bibinfo
  {author} {\bibfnamefont {D.~A.}\ \bibnamefont {Case}},\ }\bibfield  {title}
  {\bibinfo {title} {Development and testing of a general amber force field},\
  }\href@noop {} {\bibfield  {journal} {\bibinfo  {journal} {Journal of
  computational chemistry}\ }\textbf {\bibinfo {volume} {25}},\ \bibinfo
  {pages} {1157} (\bibinfo {year} {2004})}\BibitemShut {NoStop}%
\bibitem [{\citenamefont {Ponder}\ and\ \citenamefont
  {Case}(2003)}]{ponder2003force}%
  \BibitemOpen
  \bibfield  {author} {\bibinfo {author} {\bibfnamefont {J.~W.}\ \bibnamefont
  {Ponder}}\ and\ \bibinfo {author} {\bibfnamefont {D.~A.}\ \bibnamefont
  {Case}},\ }\bibfield  {title} {\bibinfo {title} {Force fields for protein
  simulations},\ }in\ \href@noop {} {\emph {\bibinfo {booktitle} {Advances in
  protein chemistry}}},\ Vol.~\bibinfo {volume} {66}\ (\bibinfo  {publisher}
  {Elsevier},\ \bibinfo {year} {2003})\ pp.\ \bibinfo {pages}
  {27--85}\BibitemShut {NoStop}%
\bibitem [{\citenamefont {Marx}\ and\ \citenamefont
  {Hutter}(2009)}]{marx2009ab}%
  \BibitemOpen
  \bibfield  {author} {\bibinfo {author} {\bibfnamefont {D.}~\bibnamefont
  {Marx}}\ and\ \bibinfo {author} {\bibfnamefont {J.}~\bibnamefont {Hutter}},\
  }\href@noop {} {\emph {\bibinfo {title} {Ab initio molecular dynamics: basic
  theory and advanced methods}}}\ (\bibinfo  {publisher} {Cambridge University
  Press},\ \bibinfo {year} {2009})\BibitemShut {NoStop}%
\bibitem [{\citenamefont {Buckingham}(1938)}]{buckingham1938classical}%
  \BibitemOpen
  \bibfield  {author} {\bibinfo {author} {\bibfnamefont {R.~A.}\ \bibnamefont
  {Buckingham}},\ }\bibfield  {title} {\bibinfo {title} {The classical equation
  of state of gaseous helium, neon and argon},\ }\href@noop {} {\bibfield
  {journal} {\bibinfo  {journal} {Proceedings of the Royal Society of London.
  Series A. Mathematical and Physical Sciences}\ }\textbf {\bibinfo {volume}
  {168}},\ \bibinfo {pages} {264} (\bibinfo {year} {1938})}\BibitemShut
  {NoStop}%
\bibitem [{\citenamefont {Hockney}\ and\ \citenamefont
  {Eastwood}(1988)}]{hockney1988computer_kspacepppm}%
  \BibitemOpen
  \bibfield  {author} {\bibinfo {author} {\bibfnamefont {R.~W.}\ \bibnamefont
  {Hockney}}\ and\ \bibinfo {author} {\bibfnamefont {J.~W.}\ \bibnamefont
  {Eastwood}},\ }\href@noop {} {\emph {\bibinfo {title} {Computer simulation
  using particles}}}\ (\bibinfo  {publisher} {crc Press},\ \bibinfo {year}
  {1988})\BibitemShut {NoStop}%
\bibitem [{\citenamefont {Frisch}\ \emph {et~al.}(2009)\citenamefont {Frisch},
  \citenamefont {Trucks}, \citenamefont {Schlegel}, \citenamefont {Scuseria},
  \citenamefont {Robb}, \citenamefont {Cheeseman}, \citenamefont {Scalmani},
  \citenamefont {Barone}, \citenamefont {Mennucci}, \citenamefont {Petersson},
  \citenamefont {Nakatsuji}, \citenamefont {Caricato}, \citenamefont {Li},
  \citenamefont {Hratchian}, \citenamefont {Izmaylov}, \citenamefont {Bloino},
  \citenamefont {Zheng}, \citenamefont {Sonnenberg}, \citenamefont {Hada},
  \citenamefont {Ehara}, \citenamefont {Toyota}, \citenamefont {Fukuda},
  \citenamefont {Hasegawa}, \citenamefont {Ishida}, \citenamefont {Nakajima},
  \citenamefont {Honda}, \citenamefont {Kitao}, \citenamefont {Nakai},
  \citenamefont {Vreven}, \citenamefont {Montgomery}, \citenamefont {Jr.},
  \citenamefont {Peralta}, \citenamefont {Ogliaro}, \citenamefont {Bearpark},
  \citenamefont {Heyd}, \citenamefont {Brothers}, \citenamefont {Kudin},
  \citenamefont {Staroverov}, \citenamefont {Kobayashi}, \citenamefont
  {Normand}, \citenamefont {Raghavachari}, \citenamefont {Rendell},
  \citenamefont {Burant}, \citenamefont {Iyengar}, \citenamefont {Tomasi},
  \citenamefont {Cossi}, \citenamefont {Rega}, \citenamefont {Millam},
  \citenamefont {Klene}, \citenamefont {Knox}, \citenamefont {Cross},
  \citenamefont {Bakken}, \citenamefont {Adamo}, \citenamefont {Jaramillo},
  \citenamefont {Gomperts}, \citenamefont {Stratmann}, \citenamefont {Yazyev},
  \citenamefont {Austin}, \citenamefont {Cammi}, \citenamefont {Pomelli},
  \citenamefont {Ochterski}, \citenamefont {Martin}, \citenamefont {Morokuma},
  \citenamefont {Zakrzewski}, \citenamefont {Voth}, \citenamefont {Salvador},
  \citenamefont {Dannenberg}, \citenamefont {Dapprich}, \citenamefont
  {Daniels}, \citenamefont {Farkas}, \citenamefont {Foresman}, \citenamefont
  {Ortiz}, \citenamefont {Cioslowski}, ,\ and\ \citenamefont {Fox}}]{g09}%
  \BibitemOpen
  \bibfield  {author} {\bibinfo {author} {\bibfnamefont {M.~J.}\ \bibnamefont
  {Frisch}}, \bibinfo {author} {\bibfnamefont {G.~W.}\ \bibnamefont {Trucks}},
  \bibinfo {author} {\bibfnamefont {H.~B.}\ \bibnamefont {Schlegel}}, \bibinfo
  {author} {\bibfnamefont {G.~E.}\ \bibnamefont {Scuseria}}, \bibinfo {author}
  {\bibfnamefont {M.~A.}\ \bibnamefont {Robb}}, \bibinfo {author}
  {\bibfnamefont {J.~R.}\ \bibnamefont {Cheeseman}}, \bibinfo {author}
  {\bibfnamefont {G.}~\bibnamefont {Scalmani}}, \bibinfo {author}
  {\bibfnamefont {V.}~\bibnamefont {Barone}}, \bibinfo {author} {\bibfnamefont
  {B.}~\bibnamefont {Mennucci}}, \bibinfo {author} {\bibfnamefont {G.~A.}\
  \bibnamefont {Petersson}}, \bibinfo {author} {\bibfnamefont {H.}~\bibnamefont
  {Nakatsuji}}, \bibinfo {author} {\bibfnamefont {M.}~\bibnamefont {Caricato}},
  \bibinfo {author} {\bibfnamefont {X.}~\bibnamefont {Li}}, \bibinfo {author}
  {\bibfnamefont {H.~P.}\ \bibnamefont {Hratchian}}, \bibinfo {author}
  {\bibfnamefont {A.~F.}\ \bibnamefont {Izmaylov}}, \bibinfo {author}
  {\bibfnamefont {J.}~\bibnamefont {Bloino}}, \bibinfo {author} {\bibfnamefont
  {G.}~\bibnamefont {Zheng}}, \bibinfo {author} {\bibfnamefont {J.~L.}\
  \bibnamefont {Sonnenberg}}, \bibinfo {author} {\bibfnamefont
  {M.}~\bibnamefont {Hada}}, \bibinfo {author} {\bibfnamefont {M.}~\bibnamefont
  {Ehara}}, \bibinfo {author} {\bibfnamefont {K.}~\bibnamefont {Toyota}},
  \bibinfo {author} {\bibfnamefont {R.}~\bibnamefont {Fukuda}}, \bibinfo
  {author} {\bibfnamefont {J.}~\bibnamefont {Hasegawa}}, \bibinfo {author}
  {\bibfnamefont {M.}~\bibnamefont {Ishida}}, \bibinfo {author} {\bibfnamefont
  {T.}~\bibnamefont {Nakajima}}, \bibinfo {author} {\bibfnamefont
  {Y.}~\bibnamefont {Honda}}, \bibinfo {author} {\bibfnamefont
  {O.}~\bibnamefont {Kitao}}, \bibinfo {author} {\bibfnamefont
  {H.}~\bibnamefont {Nakai}}, \bibinfo {author} {\bibfnamefont
  {T.}~\bibnamefont {Vreven}}, \bibinfo {author} {\bibfnamefont {J.~A.}\
  \bibnamefont {Montgomery}}, \bibinfo {author} {\bibnamefont {Jr.}}, \bibinfo
  {author} {\bibfnamefont {J.~E.}\ \bibnamefont {Peralta}}, \bibinfo {author}
  {\bibfnamefont {F.}~\bibnamefont {Ogliaro}}, \bibinfo {author} {\bibfnamefont
  {M.}~\bibnamefont {Bearpark}}, \bibinfo {author} {\bibfnamefont {J.~J.}\
  \bibnamefont {Heyd}}, \bibinfo {author} {\bibfnamefont {E.}~\bibnamefont
  {Brothers}}, \bibinfo {author} {\bibfnamefont {K.~N.}\ \bibnamefont {Kudin}},
  \bibinfo {author} {\bibfnamefont {V.~N.}\ \bibnamefont {Staroverov}},
  \bibinfo {author} {\bibfnamefont {R.}~\bibnamefont {Kobayashi}}, \bibinfo
  {author} {\bibfnamefont {J.}~\bibnamefont {Normand}}, \bibinfo {author}
  {\bibfnamefont {K.}~\bibnamefont {Raghavachari}}, \bibinfo {author}
  {\bibfnamefont {A.}~\bibnamefont {Rendell}}, \bibinfo {author} {\bibfnamefont
  {J.~C.}\ \bibnamefont {Burant}}, \bibinfo {author} {\bibfnamefont {S.~S.}\
  \bibnamefont {Iyengar}}, \bibinfo {author} {\bibfnamefont {J.}~\bibnamefont
  {Tomasi}}, \bibinfo {author} {\bibfnamefont {M.}~\bibnamefont {Cossi}},
  \bibinfo {author} {\bibfnamefont {N.}~\bibnamefont {Rega}}, \bibinfo {author}
  {\bibfnamefont {J.~M.}\ \bibnamefont {Millam}}, \bibinfo {author}
  {\bibfnamefont {M.}~\bibnamefont {Klene}}, \bibinfo {author} {\bibfnamefont
  {J.~E.}\ \bibnamefont {Knox}}, \bibinfo {author} {\bibfnamefont {J.~B.}\
  \bibnamefont {Cross}}, \bibinfo {author} {\bibfnamefont {V.}~\bibnamefont
  {Bakken}}, \bibinfo {author} {\bibfnamefont {C.}~\bibnamefont {Adamo}},
  \bibinfo {author} {\bibfnamefont {J.}~\bibnamefont {Jaramillo}}, \bibinfo
  {author} {\bibfnamefont {R.}~\bibnamefont {Gomperts}}, \bibinfo {author}
  {\bibfnamefont {R.~E.}\ \bibnamefont {Stratmann}}, \bibinfo {author}
  {\bibfnamefont {O.}~\bibnamefont {Yazyev}}, \bibinfo {author} {\bibfnamefont
  {A.~J.}\ \bibnamefont {Austin}}, \bibinfo {author} {\bibfnamefont
  {R.}~\bibnamefont {Cammi}}, \bibinfo {author} {\bibfnamefont
  {C.}~\bibnamefont {Pomelli}}, \bibinfo {author} {\bibfnamefont {J.~W.}\
  \bibnamefont {Ochterski}}, \bibinfo {author} {\bibfnamefont {R.~L.}\
  \bibnamefont {Martin}}, \bibinfo {author} {\bibfnamefont {K.}~\bibnamefont
  {Morokuma}}, \bibinfo {author} {\bibfnamefont {V.~G.}\ \bibnamefont
  {Zakrzewski}}, \bibinfo {author} {\bibfnamefont {G.~A.}\ \bibnamefont
  {Voth}}, \bibinfo {author} {\bibfnamefont {P.}~\bibnamefont {Salvador}},
  \bibinfo {author} {\bibfnamefont {J.~J.}\ \bibnamefont {Dannenberg}},
  \bibinfo {author} {\bibfnamefont {S.}~\bibnamefont {Dapprich}}, \bibinfo
  {author} {\bibfnamefont {A.~D.}\ \bibnamefont {Daniels}}, \bibinfo {author}
  {\bibfnamefont {O.}~\bibnamefont {Farkas}}, \bibinfo {author} {\bibfnamefont
  {J.~B.}\ \bibnamefont {Foresman}}, \bibinfo {author} {\bibfnamefont {J.~V.}\
  \bibnamefont {Ortiz}}, \bibinfo {author} {\bibfnamefont {J.}~\bibnamefont
  {Cioslowski}}, ,\ and\ \bibinfo {author} {\bibfnamefont {D.~J.}\ \bibnamefont
  {Fox}},\ }\href@noop {} {\bibinfo {title} {Gaussian 09 {R}evision {D}.01}}
  (\bibinfo {year} {2009}),\ \bibinfo {note} {gaussian Inc. Wallingford
  CT}\BibitemShut {NoStop}%
\bibitem [{\citenamefont {Becke}(1988)}]{becke1988density}%
  \BibitemOpen
  \bibfield  {author} {\bibinfo {author} {\bibfnamefont {A.~D.}\ \bibnamefont
  {Becke}},\ }\bibfield  {title} {\bibinfo {title} {Density-functional
  exchange-energy approximation with correct asymptotic behavior},\ }\href@noop
  {} {\bibfield  {journal} {\bibinfo  {journal} {Physical review A}\ }\textbf
  {\bibinfo {volume} {38}},\ \bibinfo {pages} {3098} (\bibinfo {year}
  {1988})}\BibitemShut {NoStop}%
\bibitem [{\citenamefont {Lee}\ \emph {et~al.}(1988)\citenamefont {Lee},
  \citenamefont {Yang},\ and\ \citenamefont {Parr}}]{lee1988development}%
  \BibitemOpen
  \bibfield  {author} {\bibinfo {author} {\bibfnamefont {C.}~\bibnamefont
  {Lee}}, \bibinfo {author} {\bibfnamefont {W.}~\bibnamefont {Yang}},\ and\
  \bibinfo {author} {\bibfnamefont {R.~G.}\ \bibnamefont {Parr}},\ }\bibfield
  {title} {\bibinfo {title} {Development of the colle-salvetti
  correlation-energy formula into a functional of the electron density},\
  }\href@noop {} {\bibfield  {journal} {\bibinfo  {journal} {Physical review
  B}\ }\textbf {\bibinfo {volume} {37}},\ \bibinfo {pages} {785} (\bibinfo
  {year} {1988})}\BibitemShut {NoStop}%
\bibitem [{\citenamefont {Vosko}\ \emph {et~al.}(1980)\citenamefont {Vosko},
  \citenamefont {Wilk},\ and\ \citenamefont {Nusair}}]{vosko1980accurate}%
  \BibitemOpen
  \bibfield  {author} {\bibinfo {author} {\bibfnamefont {S.~H.}\ \bibnamefont
  {Vosko}}, \bibinfo {author} {\bibfnamefont {L.}~\bibnamefont {Wilk}},\ and\
  \bibinfo {author} {\bibfnamefont {M.}~\bibnamefont {Nusair}},\ }\bibfield
  {title} {\bibinfo {title} {Accurate spin-dependent electron liquid
  correlation energies for local spin density calculations: a critical
  analysis},\ }\href@noop {} {\bibfield  {journal} {\bibinfo  {journal}
  {Canadian Journal of physics}\ }\textbf {\bibinfo {volume} {58}},\ \bibinfo
  {pages} {1200} (\bibinfo {year} {1980})}\BibitemShut {NoStop}%
\bibitem [{\citenamefont {Stephens}\ \emph {et~al.}(1994)\citenamefont
  {Stephens}, \citenamefont {Devlin}, \citenamefont {Chabalowski},\ and\
  \citenamefont {Frisch}}]{stephens1994ab}%
  \BibitemOpen
  \bibfield  {author} {\bibinfo {author} {\bibfnamefont {P.~J.}\ \bibnamefont
  {Stephens}}, \bibinfo {author} {\bibfnamefont {F.~J.}\ \bibnamefont
  {Devlin}}, \bibinfo {author} {\bibfnamefont {C.~F.}\ \bibnamefont
  {Chabalowski}},\ and\ \bibinfo {author} {\bibfnamefont {M.~J.}\ \bibnamefont
  {Frisch}},\ }\bibfield  {title} {\bibinfo {title} {Ab initio calculation of
  vibrational absorption and circular dichroism spectra using density
  functional force fields},\ }\href@noop {} {\bibfield  {journal} {\bibinfo
  {journal} {The Journal of physical chemistry}\ }\textbf {\bibinfo {volume}
  {98}},\ \bibinfo {pages} {11623} (\bibinfo {year} {1994})}\BibitemShut
  {NoStop}%
\bibitem [{\citenamefont {McLean}\ and\ \citenamefont
  {Chandler}(1980)}]{mclean1980contracted}%
  \BibitemOpen
  \bibfield  {author} {\bibinfo {author} {\bibfnamefont {A.}~\bibnamefont
  {McLean}}\ and\ \bibinfo {author} {\bibfnamefont {G.}~\bibnamefont
  {Chandler}},\ }\bibfield  {title} {\bibinfo {title} {Contracted gaussian
  basis sets for molecular calculations. i. second row atoms, z= 11--18},\
  }\href@noop {} {\bibfield  {journal} {\bibinfo  {journal} {J. Chem. Phys.}\
  }\textbf {\bibinfo {volume} {72}},\ \bibinfo {pages} {5639} (\bibinfo {year}
  {1980})}\BibitemShut {NoStop}%
\bibitem [{\citenamefont {Clark}\ \emph {et~al.}(1983)\citenamefont {Clark},
  \citenamefont {Chandrasekhar}, \citenamefont {Spitznagel},\ and\
  \citenamefont {Schleyer}}]{clark1983a}%
  \BibitemOpen
  \bibfield  {author} {\bibinfo {author} {\bibfnamefont {T.}~\bibnamefont
  {Clark}}, \bibinfo {author} {\bibfnamefont {J.}~\bibnamefont
  {Chandrasekhar}}, \bibinfo {author} {\bibfnamefont {G.~W.}\ \bibnamefont
  {Spitznagel}},\ and\ \bibinfo {author} {\bibfnamefont {P.~V.~R.}\
  \bibnamefont {Schleyer}},\ }\bibfield  {title} {\bibinfo {title} {Efficient
  diffuse function-augmented basis sets for anion calculations. iii. the 3-21+g
  basis set for first-row elements, li-f},\ }\bibfield  {journal} {\bibinfo
  {journal} {J. Comput. Chem.}\ }\textbf {\bibinfo {volume} {4}},\ \href
  {https://doi.org/10.1002/jcc.540040303} {10.1002/jcc.540040303} (\bibinfo
  {year} {1983})\BibitemShut {NoStop}%
\bibitem [{\citenamefont {Krishnan}\ \emph {et~al.}(1980)\citenamefont
  {Krishnan}, \citenamefont {Binkley}, \citenamefont {Seeger},\ and\
  \citenamefont {Pople}}]{krishnan1980a}%
  \BibitemOpen
  \bibfield  {author} {\bibinfo {author} {\bibfnamefont {R.}~\bibnamefont
  {Krishnan}}, \bibinfo {author} {\bibfnamefont {J.~S.}\ \bibnamefont
  {Binkley}}, \bibinfo {author} {\bibfnamefont {R.}~\bibnamefont {Seeger}},\
  and\ \bibinfo {author} {\bibfnamefont {J.~A.}\ \bibnamefont {Pople}},\
  }\bibfield  {title} {\bibinfo {title} {Self-consistent molecular orbital
  methods. xx. a basis set for correlated wave functions},\ }\bibfield
  {journal} {\bibinfo  {journal} {J. Chem. Phys.}\ }\textbf {\bibinfo {volume}
  {72}},\ \href {https://doi.org/10.1063/1.438955} {10.1063/1.438955} (\bibinfo
  {year} {1980})\BibitemShut {NoStop}%
\bibitem [{\citenamefont {Becke}(1993)}]{becke1993density}%
  \BibitemOpen
  \bibfield  {author} {\bibinfo {author} {\bibfnamefont {A.~D.}\ \bibnamefont
  {Becke}},\ }\bibfield  {title} {\bibinfo {title} {Density‐functional
  thermochemistry. iii. the role of exact exchange},\ }\href
  {https://doi.org/10.1063/1.464913} {\bibfield  {journal} {\bibinfo  {journal}
  {J. Chem. Phys.}\ }\textbf {\bibinfo {volume} {98}},\ \bibinfo {pages} {5648}
  (\bibinfo {year} {1993})}\BibitemShut {NoStop}%
\bibitem [{\citenamefont {Perdew}\ and\ \citenamefont
  {Yue}(1986)}]{perdew1986accurate}%
  \BibitemOpen
  \bibfield  {author} {\bibinfo {author} {\bibfnamefont {J.~P.}\ \bibnamefont
  {Perdew}}\ and\ \bibinfo {author} {\bibfnamefont {W.}~\bibnamefont {Yue}},\
  }\bibfield  {title} {\bibinfo {title} {Accurate and simple density functional
  for the electronic exchange energy: Generalized gradient approximation},\
  }\href@noop {} {\bibfield  {journal} {\bibinfo  {journal} {Physical review
  B}\ }\textbf {\bibinfo {volume} {33}},\ \bibinfo {pages} {8800} (\bibinfo
  {year} {1986})}\BibitemShut {NoStop}%
\bibitem [{\citenamefont {Perdew}\ and\ \citenamefont
  {Wang}(1992)}]{perdew1992accurate}%
  \BibitemOpen
  \bibfield  {author} {\bibinfo {author} {\bibfnamefont {J.~P.}\ \bibnamefont
  {Perdew}}\ and\ \bibinfo {author} {\bibfnamefont {Y.}~\bibnamefont {Wang}},\
  }\bibfield  {title} {\bibinfo {title} {Accurate and simple analytic
  representation of the electron-gas correlation energy},\ }\href@noop {}
  {\bibfield  {journal} {\bibinfo  {journal} {Physical review B}\ }\textbf
  {\bibinfo {volume} {45}},\ \bibinfo {pages} {13244} (\bibinfo {year}
  {1992})}\BibitemShut {NoStop}%
\bibitem [{\citenamefont {Weigend}\ and\ \citenamefont
  {Ahlrichs}(2005)}]{weigend2005balanced}%
  \BibitemOpen
  \bibfield  {author} {\bibinfo {author} {\bibfnamefont {F.}~\bibnamefont
  {Weigend}}\ and\ \bibinfo {author} {\bibfnamefont {R.}~\bibnamefont
  {Ahlrichs}},\ }\bibfield  {title} {\bibinfo {title} {Balanced basis sets of
  split valence, triple zeta valence and quadruple zeta valence quality for h
  to rn: Design and assessment of accuracy},\ }\href@noop {} {\bibfield
  {journal} {\bibinfo  {journal} {Physical Chemistry Chemical Physics}\
  }\textbf {\bibinfo {volume} {7}},\ \bibinfo {pages} {3297} (\bibinfo {year}
  {2005})}\BibitemShut {NoStop}%
\bibitem [{\citenamefont {Weigend}(2006)}]{weigend2006accurate}%
  \BibitemOpen
  \bibfield  {author} {\bibinfo {author} {\bibfnamefont {F.}~\bibnamefont
  {Weigend}},\ }\bibfield  {title} {\bibinfo {title} {Accurate coulomb-fitting
  basis sets for h to rn},\ }\href@noop {} {\bibfield  {journal} {\bibinfo
  {journal} {Physical chemistry chemical physics}\ }\textbf {\bibinfo {volume}
  {8}},\ \bibinfo {pages} {1057} (\bibinfo {year} {2006})}\BibitemShut
  {NoStop}%
\bibitem [{\citenamefont {Bayly}\ \emph {et~al.}(1993)\citenamefont {Bayly},
  \citenamefont {Cieplak}, \citenamefont {Cornell},\ and\ \citenamefont
  {Kollman}}]{bayly1993well}%
  \BibitemOpen
  \bibfield  {author} {\bibinfo {author} {\bibfnamefont {C.~I.}\ \bibnamefont
  {Bayly}}, \bibinfo {author} {\bibfnamefont {P.}~\bibnamefont {Cieplak}},
  \bibinfo {author} {\bibfnamefont {W.}~\bibnamefont {Cornell}},\ and\ \bibinfo
  {author} {\bibfnamefont {P.~A.}\ \bibnamefont {Kollman}},\ }\bibfield
  {title} {\bibinfo {title} {A well-behaved electrostatic potential based
  method using charge restraints for deriving atomic charges: the resp model},\
  }\href@noop {} {\bibfield  {journal} {\bibinfo  {journal} {The Journal of
  Physical Chemistry}\ }\textbf {\bibinfo {volume} {97}},\ \bibinfo {pages}
  {10269} (\bibinfo {year} {1993})}\BibitemShut {NoStop}%
\bibitem [{\citenamefont {Wang}\ \emph {et~al.}(2006)\citenamefont {Wang},
  \citenamefont {Wang}, \citenamefont {Kollman},\ and\ \citenamefont
  {Case}}]{wang2006automatic}%
  \BibitemOpen
  \bibfield  {author} {\bibinfo {author} {\bibfnamefont {J.}~\bibnamefont
  {Wang}}, \bibinfo {author} {\bibfnamefont {W.}~\bibnamefont {Wang}}, \bibinfo
  {author} {\bibfnamefont {P.~A.}\ \bibnamefont {Kollman}},\ and\ \bibinfo
  {author} {\bibfnamefont {D.~A.}\ \bibnamefont {Case}},\ }\bibfield  {title}
  {\bibinfo {title} {Automatic atom type and bond type perception in molecular
  mechanical calculations},\ }\href@noop {} {\bibfield  {journal} {\bibinfo
  {journal} {Journal of molecular graphics and modelling}\ }\textbf {\bibinfo
  {volume} {25}},\ \bibinfo {pages} {247} (\bibinfo {year} {2006})}\BibitemShut
  {NoStop}%
\bibitem [{\citenamefont {Duke}\ \emph {et~al.}(2016)\citenamefont {Duke},
  \citenamefont {Giese}, \citenamefont {Gohlke}, \citenamefont {Goetz},
  \citenamefont {Homeyer}, \citenamefont {Izadi}, \citenamefont {Janowski},
  \citenamefont {Kaus}, \citenamefont {Kovalenko}, \citenamefont {Lee} \emph
  {et~al.}}]{duke2016amber}%
  \BibitemOpen
  \bibfield  {author} {\bibinfo {author} {\bibfnamefont {R.}~\bibnamefont
  {Duke}}, \bibinfo {author} {\bibfnamefont {T.}~\bibnamefont {Giese}},
  \bibinfo {author} {\bibfnamefont {H.}~\bibnamefont {Gohlke}}, \bibinfo
  {author} {\bibfnamefont {A.}~\bibnamefont {Goetz}}, \bibinfo {author}
  {\bibfnamefont {N.}~\bibnamefont {Homeyer}}, \bibinfo {author} {\bibfnamefont
  {S.}~\bibnamefont {Izadi}}, \bibinfo {author} {\bibfnamefont
  {P.}~\bibnamefont {Janowski}}, \bibinfo {author} {\bibfnamefont
  {J.}~\bibnamefont {Kaus}}, \bibinfo {author} {\bibfnamefont {A.}~\bibnamefont
  {Kovalenko}}, \bibinfo {author} {\bibfnamefont {T.}~\bibnamefont {Lee}},
  \emph {et~al.},\ }\bibfield  {title} {\bibinfo {title} {Amber 2016},\
  }\href@noop {} {\bibfield  {journal} {\bibinfo  {journal} {University of
  California, San Francisco}\ }\textbf {\bibinfo {volume} {1}} (\bibinfo {year}
  {2016})}\BibitemShut {NoStop}%
\bibitem [{\citenamefont {Salomon-Ferrer}\ \emph {et~al.}(2013)\citenamefont
  {Salomon-Ferrer}, \citenamefont {Case},\ and\ \citenamefont
  {Walker}}]{salomon2013overview}%
  \BibitemOpen
  \bibfield  {author} {\bibinfo {author} {\bibfnamefont {R.}~\bibnamefont
  {Salomon-Ferrer}}, \bibinfo {author} {\bibfnamefont {D.~A.}\ \bibnamefont
  {Case}},\ and\ \bibinfo {author} {\bibfnamefont {R.~C.}\ \bibnamefont
  {Walker}},\ }\bibfield  {title} {\bibinfo {title} {An overview of the amber
  biomolecular simulation package},\ }\href@noop {} {\bibfield  {journal}
  {\bibinfo  {journal} {Wiley Interdisciplinary Reviews: Computational
  Molecular Science}\ }\textbf {\bibinfo {volume} {3}},\ \bibinfo {pages} {198}
  (\bibinfo {year} {2013})}\BibitemShut {NoStop}%
\bibitem [{\citenamefont {Plimpton}(1995)}]{plimpton1995fast}%
  \BibitemOpen
  \bibfield  {author} {\bibinfo {author} {\bibfnamefont {S.}~\bibnamefont
  {Plimpton}},\ }\bibfield  {title} {\bibinfo {title} {Fast parallel algorithms
  for short-range molecular dynamics},\ }\href@noop {} {\bibfield  {journal}
  {\bibinfo  {journal} {Journal of computational physics}\ }\textbf {\bibinfo
  {volume} {117}},\ \bibinfo {pages} {1} (\bibinfo {year} {1995})}\BibitemShut
  {NoStop}%
\bibitem [{\citenamefont {Larsen}\ \emph {et~al.}(2017)\citenamefont {Larsen},
  \citenamefont {Mortensen}, \citenamefont {Blomqvist}, \citenamefont
  {Castelli}, \citenamefont {Christensen}, \citenamefont {Du{\l}ak},
  \citenamefont {Friis}, \citenamefont {Groves}, \citenamefont {Hammer},
  \citenamefont {Hargus} \emph {et~al.}}]{larsen2017atomic}%
  \BibitemOpen
  \bibfield  {author} {\bibinfo {author} {\bibfnamefont {A.~H.}\ \bibnamefont
  {Larsen}}, \bibinfo {author} {\bibfnamefont {J.~J.}\ \bibnamefont
  {Mortensen}}, \bibinfo {author} {\bibfnamefont {J.}~\bibnamefont
  {Blomqvist}}, \bibinfo {author} {\bibfnamefont {I.~E.}\ \bibnamefont
  {Castelli}}, \bibinfo {author} {\bibfnamefont {R.}~\bibnamefont
  {Christensen}}, \bibinfo {author} {\bibfnamefont {M.}~\bibnamefont
  {Du{\l}ak}}, \bibinfo {author} {\bibfnamefont {J.}~\bibnamefont {Friis}},
  \bibinfo {author} {\bibfnamefont {M.~N.}\ \bibnamefont {Groves}}, \bibinfo
  {author} {\bibfnamefont {B.}~\bibnamefont {Hammer}}, \bibinfo {author}
  {\bibfnamefont {C.}~\bibnamefont {Hargus}}, \emph {et~al.},\ }\bibfield
  {title} {\bibinfo {title} {The atomic simulation environment—a python
  library for working with atoms},\ }\href@noop {} {\bibfield  {journal}
  {\bibinfo  {journal} {Journal of Physics: Condensed Matter}\ }\textbf
  {\bibinfo {volume} {29}},\ \bibinfo {pages} {273002} (\bibinfo {year}
  {2017})}\BibitemShut {NoStop}%
\bibitem [{\citenamefont {Melchionna}\ \emph {et~al.}(1993)\citenamefont
  {Melchionna}, \citenamefont {Ciccotti},\ and\ \citenamefont
  {Lee~Holian}}]{melchionna1993hoover}%
  \BibitemOpen
  \bibfield  {author} {\bibinfo {author} {\bibfnamefont {S.}~\bibnamefont
  {Melchionna}}, \bibinfo {author} {\bibfnamefont {G.}~\bibnamefont
  {Ciccotti}},\ and\ \bibinfo {author} {\bibfnamefont {B.}~\bibnamefont
  {Lee~Holian}},\ }\bibfield  {title} {\bibinfo {title} {Hoover npt dynamics
  for systems varying in shape and size},\ }\href@noop {} {\bibfield  {journal}
  {\bibinfo  {journal} {Molecular Physics}\ }\textbf {\bibinfo {volume} {78}},\
  \bibinfo {pages} {533} (\bibinfo {year} {1993})}\BibitemShut {NoStop}%
\bibitem [{\citenamefont {Melchionna}(2000)}]{melchionna2000constrained}%
  \BibitemOpen
  \bibfield  {author} {\bibinfo {author} {\bibfnamefont {S.}~\bibnamefont
  {Melchionna}},\ }\bibfield  {title} {\bibinfo {title} {Constrained systems
  and statistical distribution},\ }\href@noop {} {\bibfield  {journal}
  {\bibinfo  {journal} {Physical Review E}\ }\textbf {\bibinfo {volume} {61}},\
  \bibinfo {pages} {6165} (\bibinfo {year} {2000})}\BibitemShut {NoStop}%
\bibitem [{\citenamefont {Holian}\ \emph {et~al.}(1990)\citenamefont {Holian},
  \citenamefont {De~Groot}, \citenamefont {Hoover},\ and\ \citenamefont
  {Hoover}}]{holian1990time}%
  \BibitemOpen
  \bibfield  {author} {\bibinfo {author} {\bibfnamefont {B.~L.}\ \bibnamefont
  {Holian}}, \bibinfo {author} {\bibfnamefont {A.~J.}\ \bibnamefont
  {De~Groot}}, \bibinfo {author} {\bibfnamefont {W.~G.}\ \bibnamefont
  {Hoover}},\ and\ \bibinfo {author} {\bibfnamefont {C.~G.}\ \bibnamefont
  {Hoover}},\ }\bibfield  {title} {\bibinfo {title} {Time-reversible
  equilibrium and nonequilibrium isothermal-isobaric simulations with
  centered-difference stoermer algorithms},\ }\href@noop {} {\bibfield
  {journal} {\bibinfo  {journal} {Physical Review A}\ }\textbf {\bibinfo
  {volume} {41}},\ \bibinfo {pages} {4552} (\bibinfo {year}
  {1990})}\BibitemShut {NoStop}%
\bibitem [{\citenamefont {Oganov}\ and\ \citenamefont
  {Valle}(2009)}]{oganov2009quantify}%
  \BibitemOpen
  \bibfield  {author} {\bibinfo {author} {\bibfnamefont {A.~R.}\ \bibnamefont
  {Oganov}}\ and\ \bibinfo {author} {\bibfnamefont {M.}~\bibnamefont {Valle}},\
  }\bibfield  {title} {\bibinfo {title} {How to quantify energy landscapes of
  solids},\ }\href@noop {} {\bibfield  {journal} {\bibinfo  {journal} {The
  Journal of chemical physics}\ }\textbf {\bibinfo {volume} {130}},\ \bibinfo
  {pages} {104504} (\bibinfo {year} {2009})}\BibitemShut {NoStop}%
\bibitem [{\citenamefont {Lyakhov}\ \emph {et~al.}(2010)\citenamefont
  {Lyakhov}, \citenamefont {Oganov},\ and\ \citenamefont
  {Valle}}]{lyakhov2010predict}%
  \BibitemOpen
  \bibfield  {author} {\bibinfo {author} {\bibfnamefont {A.~O.}\ \bibnamefont
  {Lyakhov}}, \bibinfo {author} {\bibfnamefont {A.~R.}\ \bibnamefont
  {Oganov}},\ and\ \bibinfo {author} {\bibfnamefont {M.}~\bibnamefont
  {Valle}},\ }\bibfield  {title} {\bibinfo {title} {How to predict very large
  and complex crystal structures},\ }\href@noop {} {\bibfield  {journal}
  {\bibinfo  {journal} {Computer Physics Communications}\ }\textbf {\bibinfo
  {volume} {181}},\ \bibinfo {pages} {1623} (\bibinfo {year}
  {2010})}\BibitemShut {NoStop}%
\bibitem [{\citenamefont {Giannozzi}\ \emph {et~al.}(2009)\citenamefont
  {Giannozzi}, \citenamefont {Baroni}, \citenamefont {Bonini}, \citenamefont
  {Calandra}, \citenamefont {Car}, \citenamefont {Cavazzoni}, \citenamefont
  {Ceresoli}, \citenamefont {Chiarotti}, \citenamefont {Cococcioni},
  \citenamefont {Dabo} \emph {et~al.}}]{giannozzi2009quantum}%
  \BibitemOpen
  \bibfield  {author} {\bibinfo {author} {\bibfnamefont {P.}~\bibnamefont
  {Giannozzi}}, \bibinfo {author} {\bibfnamefont {S.}~\bibnamefont {Baroni}},
  \bibinfo {author} {\bibfnamefont {N.}~\bibnamefont {Bonini}}, \bibinfo
  {author} {\bibfnamefont {M.}~\bibnamefont {Calandra}}, \bibinfo {author}
  {\bibfnamefont {R.}~\bibnamefont {Car}}, \bibinfo {author} {\bibfnamefont
  {C.}~\bibnamefont {Cavazzoni}}, \bibinfo {author} {\bibfnamefont
  {D.}~\bibnamefont {Ceresoli}}, \bibinfo {author} {\bibfnamefont {G.~L.}\
  \bibnamefont {Chiarotti}}, \bibinfo {author} {\bibfnamefont {M.}~\bibnamefont
  {Cococcioni}}, \bibinfo {author} {\bibfnamefont {I.}~\bibnamefont {Dabo}},
  \emph {et~al.},\ }\bibfield  {title} {\bibinfo {title} {Quantum espresso: a
  modular and open-source software project for quantum simulations of
  materials},\ }\href@noop {} {\bibfield  {journal} {\bibinfo  {journal}
  {Journal of physics: Condensed matter}\ }\textbf {\bibinfo {volume} {21}},\
  \bibinfo {pages} {395502} (\bibinfo {year} {2009})}\BibitemShut {NoStop}%
\bibitem [{\citenamefont {Giannozzi}\ \emph {et~al.}(2017)\citenamefont
  {Giannozzi}, \citenamefont {Andreussi}, \citenamefont {Brumme}, \citenamefont
  {Bunau}, \citenamefont {Nardelli}, \citenamefont {Calandra}, \citenamefont
  {Car}, \citenamefont {Cavazzoni}, \citenamefont {Ceresoli}, \citenamefont
  {Cococcioni} \emph {et~al.}}]{giannozzi2017advanced}%
  \BibitemOpen
  \bibfield  {author} {\bibinfo {author} {\bibfnamefont {P.}~\bibnamefont
  {Giannozzi}}, \bibinfo {author} {\bibfnamefont {O.}~\bibnamefont
  {Andreussi}}, \bibinfo {author} {\bibfnamefont {T.}~\bibnamefont {Brumme}},
  \bibinfo {author} {\bibfnamefont {O.}~\bibnamefont {Bunau}}, \bibinfo
  {author} {\bibfnamefont {M.~B.}\ \bibnamefont {Nardelli}}, \bibinfo {author}
  {\bibfnamefont {M.}~\bibnamefont {Calandra}}, \bibinfo {author}
  {\bibfnamefont {R.}~\bibnamefont {Car}}, \bibinfo {author} {\bibfnamefont
  {C.}~\bibnamefont {Cavazzoni}}, \bibinfo {author} {\bibfnamefont
  {D.}~\bibnamefont {Ceresoli}}, \bibinfo {author} {\bibfnamefont
  {M.}~\bibnamefont {Cococcioni}}, \emph {et~al.},\ }\bibfield  {title}
  {\bibinfo {title} {Advanced capabilities for materials modelling with quantum
  espresso},\ }\href@noop {} {\bibfield  {journal} {\bibinfo  {journal}
  {Journal of physics: Condensed matter}\ }\textbf {\bibinfo {volume} {29}},\
  \bibinfo {pages} {465901} (\bibinfo {year} {2017})}\BibitemShut {NoStop}%
\bibitem [{\citenamefont {Garrity}\ \emph {et~al.}(2014)\citenamefont
  {Garrity}, \citenamefont {Bennett}, \citenamefont {Rabe},\ and\ \citenamefont
  {Vanderbilt}}]{garrity2014pseudopotentials}%
  \BibitemOpen
  \bibfield  {author} {\bibinfo {author} {\bibfnamefont {K.~F.}\ \bibnamefont
  {Garrity}}, \bibinfo {author} {\bibfnamefont {J.~W.}\ \bibnamefont
  {Bennett}}, \bibinfo {author} {\bibfnamefont {K.~M.}\ \bibnamefont {Rabe}},\
  and\ \bibinfo {author} {\bibfnamefont {D.}~\bibnamefont {Vanderbilt}},\
  }\bibfield  {title} {\bibinfo {title} {Pseudopotentials for high-throughput
  dft calculations},\ }\href@noop {} {\bibfield  {journal} {\bibinfo  {journal}
  {Computational Materials Science}\ }\textbf {\bibinfo {volume} {81}},\
  \bibinfo {pages} {446} (\bibinfo {year} {2014})}\BibitemShut {NoStop}%
\bibitem [{\citenamefont {Berland}\ \emph {et~al.}(2014)\citenamefont
  {Berland}, \citenamefont {Arter}, \citenamefont {Cooper}, \citenamefont
  {Lee}, \citenamefont {Lundqvist}, \citenamefont {Schroder}, \citenamefont
  {Thonhauser},\ and\ \citenamefont {Hyldgaard}}]{hyldgaard2015van}%
  \BibitemOpen
  \bibfield  {author} {\bibinfo {author} {\bibfnamefont {K.}~\bibnamefont
  {Berland}}, \bibinfo {author} {\bibfnamefont {C.~A.}\ \bibnamefont {Arter}},
  \bibinfo {author} {\bibfnamefont {V.~R.}\ \bibnamefont {Cooper}}, \bibinfo
  {author} {\bibfnamefont {K.}~\bibnamefont {Lee}}, \bibinfo {author}
  {\bibfnamefont {B.~I.}\ \bibnamefont {Lundqvist}}, \bibinfo {author}
  {\bibfnamefont {E.}~\bibnamefont {Schroder}}, \bibinfo {author}
  {\bibfnamefont {T.}~\bibnamefont {Thonhauser}},\ and\ \bibinfo {author}
  {\bibfnamefont {P.}~\bibnamefont {Hyldgaard}},\ }\bibfield  {title} {\bibinfo
  {title} {van der waals density functionals built upon the electron-gas
  tradition: Facing the challenge of competing interactions},\ }\href
  {https://doi.org/10.1063/1.4871731} {\bibfield  {journal} {\bibinfo
  {journal} {The Journal of Chemical Physics}\ }\textbf {\bibinfo {volume}
  {140}},\ \bibinfo {pages} {18A539} (\bibinfo {year} {2014})}\BibitemShut
  {NoStop}%
\bibitem [{\citenamefont {Monkhorst}\ and\ \citenamefont
  {Pack}(1976)}]{monkhorst1976special}%
  \BibitemOpen
  \bibfield  {author} {\bibinfo {author} {\bibfnamefont {H.~J.}\ \bibnamefont
  {Monkhorst}}\ and\ \bibinfo {author} {\bibfnamefont {J.~D.}\ \bibnamefont
  {Pack}},\ }\bibfield  {title} {\bibinfo {title} {Special points for
  brillouin-zone integrations},\ }\href@noop {} {\bibfield  {journal} {\bibinfo
   {journal} {Physical review B}\ }\textbf {\bibinfo {volume} {13}},\ \bibinfo
  {pages} {5188} (\bibinfo {year} {1976})}\BibitemShut {NoStop}%
\bibitem [{\citenamefont {de~Gelder}\ \emph {et~al.}(2001)\citenamefont
  {de~Gelder}, \citenamefont {Wehrens},\ and\ \citenamefont
  {Hageman}}]{de2001generalized}%
  \BibitemOpen
  \bibfield  {author} {\bibinfo {author} {\bibfnamefont {R.}~\bibnamefont
  {de~Gelder}}, \bibinfo {author} {\bibfnamefont {R.}~\bibnamefont {Wehrens}},\
  and\ \bibinfo {author} {\bibfnamefont {J.~A.}\ \bibnamefont {Hageman}},\
  }\bibfield  {title} {\bibinfo {title} {A generalized expression for the
  similarity of spectra: application to powder diffraction pattern
  classification},\ }\href@noop {} {\bibfield  {journal} {\bibinfo  {journal}
  {Journal of Computational Chemistry}\ }\textbf {\bibinfo {volume} {22}},\
  \bibinfo {pages} {273} (\bibinfo {year} {2001})}\BibitemShut {NoStop}%
\bibitem [{\citenamefont {Habermehl}\ \emph {et~al.}(2014)\citenamefont
  {Habermehl}, \citenamefont {M{\"o}rschel}, \citenamefont {Eisenbrandt},
  \citenamefont {Hammer},\ and\ \citenamefont
  {Schmidt}}]{habermehl2014structure}%
  \BibitemOpen
  \bibfield  {author} {\bibinfo {author} {\bibfnamefont {S.}~\bibnamefont
  {Habermehl}}, \bibinfo {author} {\bibfnamefont {P.}~\bibnamefont
  {M{\"o}rschel}}, \bibinfo {author} {\bibfnamefont {P.}~\bibnamefont
  {Eisenbrandt}}, \bibinfo {author} {\bibfnamefont {S.~M.}\ \bibnamefont
  {Hammer}},\ and\ \bibinfo {author} {\bibfnamefont {M.~U.}\ \bibnamefont
  {Schmidt}},\ }\bibfield  {title} {\bibinfo {title} {Structure determination
  from powder data without prior indexing, using a similarity measure based on
  cross-correlation functions},\ }\href@noop {} {\bibfield  {journal} {\bibinfo
   {journal} {Acta Crystallographica Section B: Structural Science, Crystal
  Engineering and Materials}\ }\textbf {\bibinfo {volume} {70}},\ \bibinfo
  {pages} {347} (\bibinfo {year} {2014})}\BibitemShut {NoStop}%
\bibitem [{\citenamefont {Fredericks}\ \emph {et~al.}(2021)\citenamefont
  {Fredericks}, \citenamefont {Parrish}, \citenamefont {Sayre},\ and\
  \citenamefont {Zhu}}]{fredericks2021pyxtal}%
  \BibitemOpen
  \bibfield  {author} {\bibinfo {author} {\bibfnamefont {S.}~\bibnamefont
  {Fredericks}}, \bibinfo {author} {\bibfnamefont {K.}~\bibnamefont {Parrish}},
  \bibinfo {author} {\bibfnamefont {D.}~\bibnamefont {Sayre}},\ and\ \bibinfo
  {author} {\bibfnamefont {Q.}~\bibnamefont {Zhu}},\ }\bibfield  {title}
  {\bibinfo {title} {Pyxtal: A python library for crystal structure generation
  and symmetry analysis},\ }\href@noop {} {\bibfield  {journal} {\bibinfo
  {journal} {Computer Physics Communications}\ }\textbf {\bibinfo {volume}
  {261}},\ \bibinfo {pages} {107810} (\bibinfo {year} {2021})}\BibitemShut
  {NoStop}%
\bibitem [{\citenamefont {Gong}\ \emph {et~al.}(2018)\citenamefont {Gong},
  \citenamefont {Voznyy}, \citenamefont {Jain}, \citenamefont {Liu},
  \citenamefont {Sabatini}, \citenamefont {Piontkowski}, \citenamefont
  {Walters}, \citenamefont {Bappi}, \citenamefont {Nokhrin}, \citenamefont
  {Bushuyev} \emph {et~al.}}]{gong2018electron}%
  \BibitemOpen
  \bibfield  {author} {\bibinfo {author} {\bibfnamefont {X.}~\bibnamefont
  {Gong}}, \bibinfo {author} {\bibfnamefont {O.}~\bibnamefont {Voznyy}},
  \bibinfo {author} {\bibfnamefont {A.}~\bibnamefont {Jain}}, \bibinfo {author}
  {\bibfnamefont {W.}~\bibnamefont {Liu}}, \bibinfo {author} {\bibfnamefont
  {R.}~\bibnamefont {Sabatini}}, \bibinfo {author} {\bibfnamefont
  {Z.}~\bibnamefont {Piontkowski}}, \bibinfo {author} {\bibfnamefont
  {G.}~\bibnamefont {Walters}}, \bibinfo {author} {\bibfnamefont
  {G.}~\bibnamefont {Bappi}}, \bibinfo {author} {\bibfnamefont
  {S.}~\bibnamefont {Nokhrin}}, \bibinfo {author} {\bibfnamefont
  {O.}~\bibnamefont {Bushuyev}}, \emph {et~al.},\ }\bibfield  {title} {\bibinfo
  {title} {Electron--phonon interaction in efficient perovskite blue
  emitters},\ }\href@noop {} {\bibfield  {journal} {\bibinfo  {journal} {Nature
  materials}\ }\textbf {\bibinfo {volume} {17}},\ \bibinfo {pages} {550}
  (\bibinfo {year} {2018})}\BibitemShut {NoStop}%
\bibitem [{\citenamefont {Li}\ \emph {et~al.}(2019)\citenamefont {Li},
  \citenamefont {Liu}, \citenamefont {Li}, \citenamefont {Xu}, \citenamefont
  {Wu}, \citenamefont {Han}, \citenamefont {Tao}, \citenamefont {Hong},
  \citenamefont {Luo},\ and\ \citenamefont {Sun}}]{li2019two}%
  \BibitemOpen
  \bibfield  {author} {\bibinfo {author} {\bibfnamefont {L.}~\bibnamefont
  {Li}}, \bibinfo {author} {\bibfnamefont {X.}~\bibnamefont {Liu}}, \bibinfo
  {author} {\bibfnamefont {Y.}~\bibnamefont {Li}}, \bibinfo {author}
  {\bibfnamefont {Z.}~\bibnamefont {Xu}}, \bibinfo {author} {\bibfnamefont
  {Z.}~\bibnamefont {Wu}}, \bibinfo {author} {\bibfnamefont {S.}~\bibnamefont
  {Han}}, \bibinfo {author} {\bibfnamefont {K.}~\bibnamefont {Tao}}, \bibinfo
  {author} {\bibfnamefont {M.}~\bibnamefont {Hong}}, \bibinfo {author}
  {\bibfnamefont {J.}~\bibnamefont {Luo}},\ and\ \bibinfo {author}
  {\bibfnamefont {Z.}~\bibnamefont {Sun}},\ }\bibfield  {title} {\bibinfo
  {title} {Two-dimensional hybrid perovskite-type ferroelectric for highly
  polarization-sensitive shortwave photodetection},\ }\href@noop {} {\bibfield
  {journal} {\bibinfo  {journal} {Journal of the American Chemical Society}\
  }\textbf {\bibinfo {volume} {141}},\ \bibinfo {pages} {2623} (\bibinfo {year}
  {2019})}\BibitemShut {NoStop}%
\bibitem [{\citenamefont {Mao}\ \emph {et~al.}(2018{\natexlab{b}})\citenamefont
  {Mao}, \citenamefont {Ke}, \citenamefont {Pedesseau}, \citenamefont {Wu},
  \citenamefont {Katan}, \citenamefont {Even}, \citenamefont {Wasielewski},
  \citenamefont {Stoumpos},\ and\ \citenamefont {Kanatzidis}}]{mao2018hybrid}%
  \BibitemOpen
  \bibfield  {author} {\bibinfo {author} {\bibfnamefont {L.}~\bibnamefont
  {Mao}}, \bibinfo {author} {\bibfnamefont {W.}~\bibnamefont {Ke}}, \bibinfo
  {author} {\bibfnamefont {L.}~\bibnamefont {Pedesseau}}, \bibinfo {author}
  {\bibfnamefont {Y.}~\bibnamefont {Wu}}, \bibinfo {author} {\bibfnamefont
  {C.}~\bibnamefont {Katan}}, \bibinfo {author} {\bibfnamefont
  {J.}~\bibnamefont {Even}}, \bibinfo {author} {\bibfnamefont {M.~R.}\
  \bibnamefont {Wasielewski}}, \bibinfo {author} {\bibfnamefont {C.~C.}\
  \bibnamefont {Stoumpos}},\ and\ \bibinfo {author} {\bibfnamefont {M.~G.}\
  \bibnamefont {Kanatzidis}},\ }\bibfield  {title} {\bibinfo {title} {Hybrid
  dion--jacobson 2d lead iodide perovskites},\ }\href@noop {} {\bibfield
  {journal} {\bibinfo  {journal} {Journal of the American Chemical Society}\
  }\textbf {\bibinfo {volume} {140}},\ \bibinfo {pages} {3775} (\bibinfo {year}
  {2018}{\natexlab{b}})}\BibitemShut {NoStop}%
\end{thebibliography}%

\begin{table*}[!ht]
\scriptsize
\begin{tabular}{
p{0.25\textwidth} 
>{\centering}p{0.1\textwidth} >{\centering\arraybackslash}p{0.6\textwidth}
}
\cline{2-3}
\scriptsize
&\multirow{2}{*}{Value}&\multirow{2}{*}{Description}\\
\addlinespace
\textbf{Outer loop} & & \\
\hline
\addlinespace
\multicolumn{1}{c}{\multirow{2}{*}{$\text{M}\text{H}_\text{steps}$}}&\multirow{2}{*}{170}&\multirow{2}{*}{Algorithm stops after completing $\text{M}\text{H}_\text{steps}$ minima hopping cycles}\\\\
\addlinespace
\hline
\addlinespace
\multicolumn{1}{c}{\multirow{2}{*}{$\text{md}_\text{min}$}}&\multirow{2}{*}{4000}&Number of local minima to be passed through before stopping NPT dynamics\\
\addlinespace
\hline
\addlinespace
\multicolumn{1}{c}{\multirow{2}{*}{$T_0\ /\ \text{K}$}}&\multirow{2}{*}{50}&Initial temperature of NPT dynamics; temperature is reset to $T_0$ if a minimum is accepted\\
\addlinespace
\hline
\addlinespace
\multicolumn{1}{c}{\multirow{2}{*}{$\beta$}}&\multirow{2}{*}{1.1}&Factor by which the temperature of the NPT thermostat is multiplied if a candidate minimum is rejected\\
\addlinespace
\hline
\addlinespace
\multicolumn{1}{c}{\multirow{2}{*}{$E_\text{diff}\ /\ \text{eV}$}}&\multirow{2}{*}{1.1}&If a candidate structure is greater in energy compared to the last found minimum by $E_\text{diff}$, the candidate is rejected\\
\addlinespace
\hline\\
\textbf{NPT} & & \\
\hline
\addlinespace
\multicolumn{1}{c}{\multirow{2}{*}{$\Delta t\ /\ \text{fs}$}}&\multirow{2}{*}{0.5}&\multirow{2}{*}{Timestep of NPT dynamics}\\\\
\addlinespace
\hline
\addlinespace
\multicolumn{1}{c}{\multirow{2}{*}{$P_0\ /\ \text{eV}\text{\AA}^{-3}$}}&\multirow{2}{*}{0.0}&\multirow{2}{*}{Pressure exerted on the system during NPT dynamics}\\\\
\addlinespace
\hline
\addlinespace
\multicolumn{1}{c}{\multirow{2}{*}{$\tau\ /\ \text{fs}$}}&\multirow{2}{*}{25.0}&\multirow{2}{*}{Characteristic timescale of the thermostat}\\\\
\addlinespace
\hline
\addlinespace
\multicolumn{1}{c}{\multirow{2}{*}{$W\ /\ \text{eVfs}^2\text{\AA}^{-3}$}}&\multirow{2}{*}{337.5}&\multirow{2}{*}{A constant in the barostat differential equation }\\\\
\addlinespace
\hline\\
\textbf{Local optimizations} & & \\
\hline
\addlinespace
\multicolumn{1}{c}{\multirow{2}{*}{Optimizer}}&\multirow{2}{*}{BFGS}&\multirow{2}{*}{The local optimization algorithm}\\\\
\addlinespace
\hline
\addlinespace
\multicolumn{1}{c}{\multirow{2}{*}{$F_\text{init}\ /\ \text{eV\AA}^{-1}$}}&\multirow{2}{*}{0.05}&\multirow{2}{*}{Total force convergence threshold for the initial cell optimization}\\\\
\addlinespace
\hline
\addlinespace
\multicolumn{1}{c}{\multirow{2}{*}{$F_\text{geo}\ /\ \text{eV\AA}^{-1}$}}&\multirow{2}{*}{0.1}&\multirow{2}{*}{Total force convergence threshold for geometry optimizations}\\\\
\addlinespace
\hline
\addlinespace
\multicolumn{1}{c}{\multirow{2}{*}{$F_\text{cell}\ /\ \text{eV\AA}^{-1}$}}&\multirow{2}{*}{0.01}&\multirow{2}{*}{Total force convergence threshold for cell optimizations}\\\\
\addlinespace
\hline\\
\textbf{Oganov fingerprints} & & \\
\hline
\addlinespace
\multicolumn{1}{c}{\multirow{2}{*}{$\chi^0$}}&\multirow{2}{*}{0.005}&Maximum cosine distance between two structures below which they are considered to be the same structure\\
\addlinespace
\hline
\addlinespace
\multicolumn{1}{c}{\multirow{2}{*}{$\Delta$ / \AA}}&\multirow{2}{*}{0.05}&Width of the bins into which the Oganov fingerprint components are discretized\\
\addlinespace
\hline
\addlinespace
\multicolumn{1}{c}{\multirow{2}{*}{$\sigma$ / \AA}}&\multirow{2}{*}{0.1}&\multirow{2}{*}{Standard deviation of the gaussian smearing of fingerprints}\\\\
\addlinespace
\hline
\addlinespace
\multicolumn{1}{c}{\multirow{2}{*}{$N_\sigma$}}&\multirow{2}{*}{5}&\multirow{2}{*}{Number of standard deviations $\sigma$ at which the gaussian smearing is cut off}\\\\
\addlinespace
\hline
\addlinespace
\multicolumn{1}{c}{\multirow{2}{*}{$R_c\ /$ \AA}}&\multirow{2}{*}{$l_{\text{min}}$}&Cutoff radius in Angstrom for the fingerprints. At every MH step, the shortest cell length $l_{\text{min}}$ is used\\
\addlinespace
\hline


\end{tabular}
\caption{List of parameters used in the MH algorithm.}\label{table:MH_pars}
\end{table*}

\end{document}